\documentclass[11pt]{article}
\pdfoutput=1

\newlength{\vshift}
\newlength{\hshift}
\usepackage{amssymb,amsopn}
\usepackage{slashed}
\usepackage[utf8]{inputenc}
\usepackage{amsmath}
\usepackage{graphicx}
\usepackage{color}

\renewcommand{\theequation}{\thesection.\arabic{equation}}
\newcommand{\initiate}{\setcounter{equation}{0}}

\def\d{\textrm{d}}
\def\ds{\stackrel{\star}{,}}

\def\nn{\nonumber}
\def\be{\begin{equation}}             \def\ee{\end{equation}}
\def\ba#1{\begin{array}{#1}}          \def\ea{\end{array}}
\def\bea{\begin{eqnarray} }           \def\eea{\end{eqnarray} }
\def\beann{\begin{eqnarray*} }        \def\eeann{\end{eqnarray*} }
\def\beal{\begin{eqalign}}            \def\eeal{\end{eqalign}}
\def\lab#1{\label{eq:#1}}             
\def\bsubeq{\begin{subequations}}     \def\esubeq{\end{subequations}}
\def\bitem{\begin{itemize}}           \def\eitem{\end{itemize}}

\def\m{\mu}

\begin{document}

\begin{titlepage}

\begin{center}

{\large{\bf Noncommutative Scalar Quasinormal Modes\\ 
of the
Reissner–Nordstr\" om Black Hole}}

\vspace*{1.5cm}

{{\bf Marija Dimitrijevi\' c \' Ciri\'c${}^{1}$, 
Nikola Konjik${}^{1}$ and Andjelo Samsarov${}^{2}$}}

\vspace*{1cm}

\noindent  ${}^1$ {\it Faculty of Physics, University of Belgrade}\\ {\it Studentski trg 12,
11000 Beograd, Serbia}\\

\noindent  ${}^2$ {\it Rudjer Bo\v skovi\' c Institute, Theoretical Physics Division\\
Bijeni\v cka 54, 10002 Zagreb, Croatia}\\

\end{center}

\vspace*{2cm}

\begin{abstract}

Aiming to search  for a signal of space-time noncommutativity, we  study a quasinormal mode spectrum
of the Reissner–Nordstr\" om   black hole  in the presence of a deformed space-time structure. 
In this context we study  a noncommutative (NC) deformation of  a  scalar field,  minimally coupled
to a classical  (commutative) Reissner–Nordstr\" om
background. The deformation is performed
via a particularly chosen Killing twist to ensure that the geometry remains undeformed  (commutative). An action describing a noncommutative scalar field minimally coupled to the RN geometry is manifestly invariant under the deformed $U(1)_\star$ gauge symmetry group.
We find  the quasinormal mode solutions of the equations of motion governing the matter content of
the model in some particular range of system parameters which corresponds to a near extremal  limit.
In addition, we obtain a well defined analytical condition 
which allows for a detailed numerical analysis.
Moreover, there exists a parameter range,  rather restrictive though,
which allows for obtaining a QNMs spectrum  in a closed analytic form.
We also argue  within a semiclassical approach that NC deformation does not affect the Hawking 
temperature of thermal radiation.  
\end{abstract}
\vspace*{1cm}


\vspace*{1cm}
\quad\scriptsize{eMail:
dmarija@ipb.ac.rs, konjik@ipb.ac.rs, asamsarov@irb.hr}
\vfill

\end{titlepage}

\tableofcontents

\newpage

\section{Introduction}

     The excitation of black holes may be provoked in different ways, namely it   may  appear as a
result of merging of two black  holes, as was the case with the recent LIGO experiment \cite{ligo}
that brought about the first detection of gravitational waves, or simply by infall of matter into a
supermassive black hole.  Once being perturbed, a black hole responds to a perturbation by going
through a ringdown phase during which it emits gravitational waves. The ringdown phase of a black
hole may broadely be divided into three stages, a short period of strong initial outburst of
radiation, which is then followed by a long period of damped oscillations, dominated by quasinormal
modes (QNMs). After QNMs damp out at later times, they become superseded by a power law behaviour of
the field, known as late-time tails. Among the three stages mentioned, the second stage,
characterized
by quasinormal modes, is far more important than the other two, since it significantly dominates the
perturbation signal. This makes QNMs a very  important  object  of  study, 
not only for the reason of searching for gravitational waves, but also for other reasons,
such as the study of (in)stability of black holes
\cite{rg,konzhidinstability}, the problem of black hole area quantization 
\cite{bekenstein,Hod:1998vk,Maggiore:2007nq,Kunstatter:2002pj,vagenas,medved,Wei:2009yj}, AdS/CFT
correspondence \cite{horowitz,Birmingham:2001pj} and others.

Since the second stage of the  black hole ringdown, which is  dominated by QNMs, is essentially
independent of the details of a perturbation, it appears to be particularly suited for exploring the properties of black holes.
This feature  of  QNMs, namely that they essentially do not depend on the details of a perturbation, but only on 
the parameters of a black hole,  makes them  a fairly convenient carrier of 
information on  the properties of   black holes. 
Black hole QNMs   \cite{rg,vish,press,chandra,Cardoso:2001hn,cardosoreview,rew}  also  provide key signatures of the gravitational waves. 
Moreover, the recent experimental discovery of gravitational waves including the ringdown phase
arising from black hole
mergers \cite{ligo} have opened up new possibilities for the observations of  QNMs.
This raises to a more  solid ground the idea of
the gravitational wave astronomy as being able to  probe  the primordial universe.

Besides carrying the intrinsic information about black holes, it is reasonable to assume
that QNMs may also carry information about the properties of the underlying space-time
structure. In particular, if the underlying space-time structure is deformed  in such a way that it
departs from the usual notion of smooth space-time manifold, then this
deviation should   also  in some way be  imprinted in the QNMs' spectrum. 

There are abundant theoretical arguments which point toward a necessity for deforming a space-time
manifold at higher energies. In particular, it is well known that the quantum
theory and general relativity together lead to a noncommutative (NC) description of space-time
\cite{ahluwalia,dop1,dop2}. There are many different methods that are specifically devised for
dealing with the deformed
structures of space-time. In this paper we use the method of deformation by
a Drinfeld twist. We derive the differential calculus along the lines of \cite{PMKLWbook, PL09} and
we use the Seiberg-Witten map \cite{jssw, seiberg} to derive the NC gauge theory. The choice of
twist is non-unique and there are various examples in the literature. Inspired by the
$\kappa$-Minkowski space-time \cite{LukPrvi} with commutation relations
\begin{equation}
[x^0 \ds x^j] = x^0\star x^j - x^j \star x^0 = ia x^j,\quad
[x^i \ds x^j] =0 ,  \label{AbxStarComm}
\end{equation}
with an arbitrary constant $a$, in our previous work \cite{miU1twisted, DJP} we construced the NC
$U(1)_\star$ gauge theory based on
the twist
\begin{equation}
\mathcal{F} = e^{-\frac{ia}{2}
(\partial_t\otimes x^j\partial_j-x^j\partial_j\otimes
\partial_t)},  \label{KappaTwist}
\end{equation}
with two commuting vector fields $X_1=\partial_t$ and $X_2=x^j\partial_j$
and indices $j=1,2,3$. However, the vector field $X_2$ does not belong to the Poincar\'e algebra
and thus, the deformed (twisted) space-time symmetry of   the constructed  model is the twisted $gl(1,3)$. To keep
the space-time symmetry at the level of a twisted Poincar\'e symmetry, in this work we choose
a twist of the form
\begin{equation}
\mathcal{F} = e^{-\frac{ia}{2}
\big( \partial_t\otimes (x\partial_y-y\partial_x) - (x\partial_y-y\partial_x)\otimes
\partial_t \big) }.  \label{AngTwistUvod}
\end{equation}
This twist we call "angular twist" since the vector field $x\partial_y-y\partial_x$ is the generator
of rotations around the $z$-axis. The twist (\ref{AngTwistUvod}) has additional nice properties
when applied to some special curved backgrounds as we will explain through the paper.

In order to
find experimental signatures of noncommutativity one can
construct different field theory models on NC spaces. NC correction to Standard Model of particle
physics have been intensively discussed in literature \cite{NCSM}. Various NC extensions of General
Relativity with different phenomenological consequences have also been proposed \cite{NCGR}. Our
goal in this paper is to study quasinormal modes of a NC scalar field in a fixed background of the
Reissner–Nordstr\" om (RN) black hole.  More precisely, we fix a particular NC deformation of space-time by "turning on" the twist (1.3). In string theory, this step corresponds to turning on background fluxes. Then strings propagate in a fixed background determined by these fluxes. After fixing the NC deformation in our model, we study the propagation of scalar  and $U(1)$ gauge fields in the non-propagating geometry of RN black hole. In particular, we study the QNMs spectrum of the charged, massive scalar field. Our model is semiclassical in the following sense: the RN geometry is non-dynamical and, unlike the scalar and the gauge field, it does not feel the deformation. A more general treatment, where the geometry/gravity is dynamical is a comlpex problem and it is beyond the scope of the present paper. Nevertheless, we plan to investigate this problem in our future work.

Note that throughout the paper, exception being the end of Section 6, we use the term "semiclassical" in a somehow different context. Namely, our scalar field is not quantized in the sense of quantum field theory, but it feels the effects of space-time deformation (noncommutativity). Therefore the scalar field is noncommutative. The gravitational field is also not quantized, but in addition, it does not feel the noncommutativity.

Getting back to qusinormal modes, they were first introduced by Vishveshwara \cite{vish} as  solutions to the wave
equation that are purely ingoing at the horizon and purely outgoing at far infinity\footnote{For
asymptotically anti-de Sitter geometries this definition changes a little bit, which  effects the
corresponding QNMs spectrum \cite{wang,Cardoso:2001bb,Cardoso:2001vs,bertikokotas,abdalla}.
Here, we shall be interested in geometries that are asymptotically flat.}. The boundary conditions
of pure ingoingness at the horizon and pure  outgoingness  at infinity that are imposed on the
solutions of the wave equation reflect the intrinsic  inability of  black holes to sustain stable
matter-energy configurations in the region around them. Over the years an extensive machinery was
developed to calculate QNMs for various geometries and types of perturbations. The methods
range from purely numerical \cite{leaver,nollert,zhidenkonum}, all the way through a set of various
semianalytic and semiclassical methods,
\cite{schutzwill,iyer,ferarimashwkb,simone,konoplyawkb,abdalla} 
down to the  purely analytic ones. However, there are very few cases that are amenable to an exact
analytical treatment. Among them  are the  QNMs calculated in an analytically
exact way  for the case of scalar perturbations  in the nonrotating BTZ background
\cite{Cardoso:2001hn},
as well as in the rotating BTZ background perturbed by massive scalars \cite{Birmingham:2001hc}.
Fermionic perturbations in the same background were treated in 
\cite{Dasgupta:1998jg, Das:1999pt, Birmingham:2001pj} and vector perturbations in
\cite{Birmingham:2001pj}.
The quasinormal spectrum  in the pure anti–de Sitter space-time was first found in \cite{burgess}
for a
massless scalar field   and for gravitational perturbations in
four-dimensional AdS space-time was found in \cite{cardosoanalytic} and then generalized to
arbitrary space-time dimensions in \cite{natario}. Exact analytical treatments in de Sitter space
searching for QNMs spectrum were carried out in \cite{natario} and \cite{lopez2006} in arbitrary
space-time dimensions.
The analytical expressions for the
quasinormal spectra of topological AdS black holes
were found for the massive scalar field \cite{aros} and for gravitational perturbations \cite{birmmokh}.
Moreover, the exact solutions for quasinormal modes of the Gauss-Bonnet black hole
were found in \cite{pagonzalez,grozdanov}.
Likewise, the analytical approximations for quasinormal modes of dilatonic black holes were found in \cite{becar} and \cite{lopez2009} and for  $d$-dimensional Schwarzschild black hole with Gauss-Bonnet term in \cite{chak-gupta}.

Analytic treatment may also  be possible in other settings, though for a limited ranges of system
parameters, such as the near extremal regime of black holes
\cite{cardosonearextremal, molina, Hod:2010hw},  or different asymptotic limits,
including large multipole number expansion or high overtone numbers
\cite{ferari,zhidenko,konoplya,vanzo,motl,tamaki,chak-gupta}.

QNMs spectra in certain settings have been  analysed in the presence of the noncommutative (NC)
structure of space-time \cite{Giri:2006rc, Gupta:2015uga, Gupta:2017lwk}.
As we have already mentioned, in this paper we continue along these lines and consider quasinormal modes
of a NC scalar field perturbing a
black hole background of the  Reissner–Nordstr\" om type.

To start with, in Section 2 we introduce the NC space-time and the NC differential calculus. More details
on the twist formalism and the twisted differential geometry are presented in the Appendix. Using
the
twisted differential calculus, in Section 3 we construct the NC $U(1)_\star$ gauge theory
coupled to a charged scalar field. The propagating fields are the charged NC scalar field and
the $U(1)_\star$ gauge field, while the gravitational field, describing the background is fixed.
The background is arbitrary with one constraint: the vector fields entering the definition of
the angular twist have to be the Killing vector fields for the chosen background. Using the twisted
differential geometry, we write the action for our model, such that it is invariant under the NC
$U(1)_\star$ gauge transformations. Then we use the Seiberg-Witten map \cite{seiberg} to expand the
NC fields in terms of the corresponding commutative fields and obtain the expanded action. The
action and the corresponding equations of motion are expanded up to first order in the NC parameter
$a$. In Section 4 we choose the background to be the Reissner–Nordstr\" om black hole. In that way
we also fix the $U(1)$ gauge field to be the electromagnetic field of the Reissner–Nordstr\" om
black hole. The scalar field remains the only propagating field and we calculate its equation of
motion in this particular background. This equation is our starting point for discussing QNMs
solutions in Section 5. We do the QNMs analysis in the near extremal region, $(r_+ -r_-)/r_+=\tau\to 0$.
This approximation enables us to find a condition for QNMs frequences analytically. However, the
condition itself has to be solved numerically. We present our solutions at the end of Section 5 and
in Section 6. In addition, in a very special region of parameters the analytic solution of the QNMs
condition can be calculated and we present it in Section 6. Both numerical and analytic solutions
show that there is a Zeeman-like splitting of the QNMs frequences for a fixed angular momentum
number $l$. Solutions depend on the magnetic number $m$ in such a way that $m$ always couples to
the NC parameter $a$. This property we discuss in Section 6. Finally, for completeness we also
calculate the changes in the Hawking temperature of this system (RN black hole and the NC scalar
field). We find no changes. We end Section 6 with some plans for future research. 


\section{Noncommutative space-time form the angular twist}

Reissner–Nordstr\" om (RN) black hole is a well known solution of Einstein equations. It
represents a charged non-rotating black hole and it is given by
\begin{equation}
{\rm d}s^2 = (1-\frac{2MG}{r}+\frac{Q^2G}{r^2}){\rm d}t^2 - \frac{{\rm
d}r^2}{1-\frac{2MG}{r}+\frac{Q^2G}{r^2}} - r^2({\rm d}\theta^2 + \sin^2\theta{\rm d}\varphi)
. \label{dsRN}
\end{equation}
Here $M$ is the mass of the RN black hole,
while $Q$ is the charge of the RN black hole. This space-time is static and spherically symmetric,
therefore it has four Killing vectors. Two of them are $\xi_1=\partial_t$ and
$\xi_2=\partial_\varphi = x\partial_y - y\partial_x$. 

We have already mentioned in Introduction that the semiclassical approach used in this paper is based on the assumption that the geometry
(gravitational field) is classical (it is not deformed by noncommutativity), while the scalar field
propagating in the RN background "feels" the effects of space-time noncommutativity,  but it is not quantized in a sense of quantum field theory. In order to
realize this approach, we choose a Killing twist. The twist is given by
\begin{equation}
\mathcal{F}=e^{-\frac{i}{2}\theta ^{AB}X_{A}\otimes X_{B}}.  \label{AngTwist}
\end{equation}
Here $\theta ^{AB}$ is a constant antisymmetric matrix
\begin{equation}
\theta^{AB} =\left( {
\begin{array}{cc}
0 & a \\ 
-a & 0
\end{array}
} \right) ,  \notag
\end{equation}
with an arbitrary constant $a$.
Indices $A,B=1,2$, while $X_{1}=\partial _t$,
$X_{2}= x\partial_y - y\partial_x$ are commuting vector fields,
$[X_{1},X_{2}]=0$. This twist fulfils the requirements
(\ref{Twcond1})-(\ref{Twcond3}).
We call (\ref{AngTwist}) "angular twist" becuase the vector field $X_{2}= x\partial _y -
y\partial_x$
is nothing else but a generator of rotations around $z$-axis, that is $X_2 =
\partial_\varphi$. The vector fields $X_1$ and $X_2$ are two Killing vectors for the metric
(\ref{dsRN}) and that is why the twist is called a Killing twist. In particular, the twist
(\ref{dsRN}) does not act on the RN metric and it does not act on the functions of the RN metric.
In this way we ensure that the geometry remains undeformed.
Let us mention that the deformation of this type is a special case of deformations introduced in \cite{Lukierski:1993hk,Lukierski:2005fc}. Similar type of twist operators that lead to a Lie algebra-type of  deformation of Minkowski space-time were also considered in \cite{Meljanac:2017oek}.

Since the twist (\ref{AngTwist}) introduces space-time noncommutativity, see (\ref{xStarx}), one might be concerned with a formulation of unitary quantum field theory \cite{NonUnitaryNCQFT}. However, it was shown in \cite{UnitaryNCQFT} that a careful formulation of quantum field theory with space-time noncommutativity leads to no problems with unitarity and causality.

Before we proceed with the construction of differential calculus, let us comment on the 
relation between the $\kappa$-Minkowski deformation introduced in \cite{miU1twisted, DJP} and the
twist we use in this paper. In the case of $\kappa$-Minkowski twist in \cite{miU1twisted, DJP}, the
vector fields defining the twist
were choosen to be $X_1=\partial_t$ and $X_2=x^j\partial_j$. The vector field $X_2=x^j\partial_j$ does not
belong to the Poincar\'e algebra, but instead to the algebra of general linear
transformations $gl(1,3)$. Therefore, the twisted symmetry of the $\kappa$-Minkowski space-time is
the twisted general linear symmetry. In the case of (\ref{AngTwist}) the vector fields defining the
twist belong to the Poincar\'e algebra and the twisted symmetry of a NC Minkowski space-time
obtained for the twist (\ref{AngTwist}) is the twisted Poincar\'e symmetry. We will not be concerned
with the deformation of Minkowski space-time in this paper. Just for the completeness, we write the
twisted Poincar\'e Hopf algebra following from (\ref{AngTwist}). We use hermitean generators. In
coordinate representation they are: $P_\mu = -i\partial_\mu$ and
$M_{\mu\nu}= i(x_\mu\partial_\nu -x_\nu\partial_\mu)$ and $\eta_{\mu\nu}= diag (+,-,-,-)$. The twisted
Poincar\'e Hopf algebra is
{\small
\begin{eqnarray}
&& \lbrack P_{\mu },P_{\nu }] = 0,\quad \lbrack M_{\mu \nu },P_{\rho }]= i(\eta
_{\nu \rho }P_{\mu }-\eta _{\mu \rho }P_{\nu }),  \nn \\
&& \lbrack M_{\mu \nu },M_{\rho \sigma }] = i(\eta _{\mu \sigma }M_{\nu \rho
}+\eta _{\nu \rho }M_{\mu \sigma }-\eta _{\mu \rho }M_{\nu \sigma }-\eta
_{\nu \sigma }M_{\mu \rho }),  \label{TwistedPoincareAlg}
\end{eqnarray}
\begin{eqnarray}
&&\Delta^{\cal F} P_{0} = P_{0}\otimes 1 + 1\otimes P_{0},\nn\\
&&\Delta^{\cal F} P_{3} = P_{3}\otimes 1 + 1\otimes P_{3},\nn\\
&&\Delta^{\cal F} P_{1} = P_{1}\otimes \cos\big( \frac{a}{2}P_0 \big) + \cos\big(
\frac{a}{2}P_0 \big)\otimes P_{1} + P_{2}\otimes \sin\big( \frac{a}{2}P_0 \big) - \sin\big(
\frac{a}{2}P_0
\big)\otimes P_{2}, \label{TwistedPoincareCoproduct}\\
&&\Delta^{\cal F} P_{2} = P_{2}\otimes \cos\big( \frac{a}{2}P_0 \big) + \cos\big(
\frac{a}{2}P_0 \big)\otimes P_{2} - P_{1}\otimes \sin\big( \frac{a}{2}P_0 \big) + \sin\big(
\frac{a}{2}P_0
\big)\otimes P_{1}, \nn
\end{eqnarray}
\begin{eqnarray}
&&\Delta^{\cal F} M_{01} = M_{01}\otimes \cos\big( \frac{a}{2}P_0 \big) + \cos\big( \frac{a}{2}P_0
\big)\otimes M_{01} + M_{02}\otimes \sin\big( \frac{a}{2}P_0 \big) - \sin\big( \frac{a}{2}P_0
\big)\otimes M_{02}\nn\\
&& \hspace*{1.5cm} -P_1\otimes\frac{a}{2}M_{12}\cos\big( \frac{a}{2}P_0 \big) +
\frac{a}{2}M_{12}\cos\big( \frac{a}{2}P_0 \big)\otimes P_1 \nn\\
&& \hspace*{1.5cm} -P_2\otimes\frac{a}{2}M_{12}\sin\big( \frac{a}{2}P_0 \big) -
\frac{a}{2}M_{12}\sin\big( \frac{a}{2}P_0 \big)\otimes P_2,\nn\\
&&\Delta^{\cal F} M_{02} = M_{02}\otimes \cos\big( \frac{a}{2}P_0 \big) + \cos\big( \frac{a}{2}P_0
\big)\otimes M_{02} - M_{01}\otimes \sin\big( \frac{a}{2}P_0 \big) + \sin\big( \frac{a}{2}P_0
\big)\otimes M_{01}\nn\\
&& \hspace*{1.5cm} -P_2\otimes\frac{a}{2}M_{12}\cos\big( \frac{a}{2}P_0 \big) +
\frac{a}{2}M_{12}\cos\big( \frac{a}{2}P_0 \big)\otimes P_2\nn\\
&& \hspace*{1.5cm} +P_1\otimes\frac{a}{2}M_{12}\sin\big( \frac{a}{2}P_0 \big) +
\frac{a}{2}M_{12}\sin\big( \frac{a}{2}P_0 \big)\otimes P_1,\nn\\
&&\Delta^{\cal F} M_{03}=M_{03}\otimes 1 + 1\otimes M_{03} -\frac{a}{2}P_3\otimes M_{12}
+\frac{a}{2}M_{12}\otimes P_3 ,\nn
\end{eqnarray}
\begin{eqnarray}
&&\Delta^{\cal F} M_{12}=M_{12}\otimes 1+1\otimes M_{12},\nn\\
&&\Delta^{\cal F} M_{13} = M_{13}\otimes \cos\big( \frac{a}{2}P_0 \big) + \cos\big( \frac{a}{2}P_0
\big)\otimes M_{13} + M_{23}\otimes \sin\big( \frac{a}{2}P_0 \big) - \sin\big( \frac{a}{2}P_0
\big)\otimes M_{23}\nn\\
&&\Delta^{\cal F} M_{23} = M_{23}\otimes \cos\big( \frac{a}{2}P_0 \big) + \cos\big( \frac{a}{2}P_0
\big)\otimes M_{23} - M_{13}\otimes \sin\big( \frac{a}{2}P_0 \big) + \sin\big( \frac{a}{2}P_0
\big)\otimes M_{13}\nn
\end{eqnarray}
\begin{eqnarray}
&& \varepsilon^{\cal F} (P_{\mu })=0, \quad \varepsilon^{\cal F} (M_{\mu \nu
})=0,\label{TwistedPoincareCounit}\\
&&S^{\cal F}(P_{\mu })=-P_{\mu },\quad S^{\cal F}(M_{\mu \nu })=-M_{\mu \nu
}.\label{TwistedPoincareAntipod}
\end{eqnarray}
}


\subsection{Twisted differential calculus}

Let us analyse consequences of the twist (\ref{AngTwist}) on the differential calculus. In this
paper we shall just use the
well known results in twisted differential geometry. More details on the twisted differential
calculus can be found in Appendix.

The $\star$-product of functions is given by
\begin{eqnarray}
f\star g &=& \mu \{ e^{\frac{ia}{2} (\partial_t\otimes
(x\partial _y - y\partial_x) - (x\partial _y - y\partial_x)\otimes \partial_t)}
f\otimes g \}\nn\\
&=& fg + \frac{ia}{2}
(\partial_t f(x\partial_yg - y\partial_x g) - \partial_t g(x\partial_y f-y\partial_x f)) + 
\mathcal{O}(a^2) . \label{fStarg}
\end{eqnarray}
This $\star $-product is noncommutative, associative and in the limit $a\rightarrow 0$ it
reduces to the usual point-wise multiplication; the
last property is guaranteed by (\ref{Twcond3}) and the associativity is
guaranteed by (\ref{Twcond1}). In this way we obtain the noncommutative algebra
of functions, i.e. the noncommutative space-time.

In the special case of coordinates, the $\star$-commutation relations are
\begin{eqnarray}
[ t \ds x ] &=& -ia y,\nonumber \\
\lbrack t \ds y \rbrack &=& iax, \label{xStarx}
\end{eqnarray}
while all other coordinates commute. These commutation relations resemble the
$\kappa$-Minkowski space-time commutation relations, that is they are linear in coordinates.

Using the $\star$-product between functions and one-forms defined in (\ref{FunctionStarForms}) we
obtain
\begin{eqnarray}
\d t\star f &=& f\star \d t, \nonumber \\
\d z\star f &=& f\star \d z, \nonumber \\ 
\d x\star f &=& \cos(ia\partial_t)f\star \d x + \sin(ia\partial_t)f\star \d y,\nn\\
\d y\star f &=& \cos(ia\partial_t)f\star \d y - \sin(ia\partial_t)f\star \d x \label{fStardx} 
\end{eqnarray}
for the $\star$-product of functions with basis one-forms.
We used that ${\cal L}_{X_2}(\mathrm{d}x)=-\mathrm{d}y$ and ${\cal
L}_{X_2}(\d y)=\d x$.

In the similar way, the wedge product between forms is deformed to the $\star$-wedge product
(\ref{WedgeStar})
\begin{equation}
\omega\wedge_\star\omega^{\prime }= \bar{\mathrm{f}}^\alpha(\omega) \wedge 
\bar{\mathrm{f}}_\alpha(\omega^{\prime }).  \nn
\end{equation}
In the special case of basis 1-forms and (\ref{AngTwist}) the wedge product is undeformed, that is
\begin{equation}
\d x^\mu\wedge_\star \d x^\nu = \d x^\mu\wedge \d x^\nu .\label{dxStardx}
\end{equation}
Since basis 1-forms anticommute the volume form remains undeformed 
\begin{equation}
\d t\wedge_\star \d x\wedge_\star \d y\wedge_\star \d z = \d t\wedge \d
x\wedge \d y\wedge \d z = \d^4x. \label{VolForm}
\end{equation}

The exterior derivative is the undeformed exterior derivative, see (\ref{Differential}). The
$\star$-derivatives following from (\ref{AngTwist}) are given by
\begin{eqnarray}
\partial^\star _t f &=& \partial_t f,\nonumber \\
\partial^\star _z f &=& \partial_z f,\nonumber \\
\partial^\star _x f &=& \cos(i\frac{a}{2}\partial_t)\partial_x f -
\sin(i\frac{a}{2}\partial_t)\partial_y f, \nn\\
\partial^\star _y f &=& \cos(i\frac{a}{2}\partial_t)\partial_y f +
\sin(i\frac{a}{2}\partial_t)\partial_x f. \label{partialStar}
\end{eqnarray}

Since the twist (\ref{AngTwist}) is an Abelian twist, the cyclicity of integral holds
\begin{equation}
\int \omega _{1}\wedge _{\star }\dots \wedge _{\star }\omega
_{p}=(-1)^{d_{1}\cdot d_{2}\dots \cdot d_{p}}\int \omega _{p}\wedge _{\star
}\omega _{1}\wedge _{\star }\dots \wedge _{\star }\omega _{p-1},  \nn
\end{equation}
with $d_{1}+d_{2}+\dots +d_{p}=4$. It can be shown that the twist (\ref{AngTwist}) fulfils an even
stronger requirement. Namely,
one can chek that the cyclicyty holds also on $\star$-products of functions
\begin{equation}
\int \d^ 4x\, f\star g  = \int \d^ 4x\, g\star f = \int \d^ 4x\, f\cdot g .\label{IntCyclfg}
\end{equation}


\subsection{Hodge dual}

In general, it is difficult to generalize the Hodge dual operation to NC spaces and NC gauge
theories. For detailed discussion see \cite{PLGaugeGravity, DJP}. However, for the twist
(\ref{AngTwist}) the Hodge dual problem has a simple solution.

Let us first analyze the problem in the Minkowski space-time twisted by (\ref{AngTwist}). Then we
can define the NC Hodge dual of a 2-form $F=\frac{1}{2}F_{\mu\nu}\star\d x^\mu\wedge_\star
\d x^\nu$ in a "natural way"
\begin{eqnarray}
*_HF = *_H(\frac{1}{2}F_{\mu\nu}\star \d x^\mu\wedge_\star
\d x^\nu) &=& \frac{1}{2}F_{\mu\nu}\star
(\epsilon^{\mu\nu\rho\sigma}\eta_{\rho\alpha}\eta_{\sigma\beta}
\d x^\alpha\wedge_\star \d x^\beta)\nn\\
&=& \frac{1}{2}\epsilon^{\mu\nu\rho\sigma}\eta_{\rho\alpha}\eta_{\sigma\beta} (F_{\mu\nu}\star
\d x^\alpha\wedge_\star \d x^\beta) .\label{HodgeF}
\end{eqnarray}
Let us further assume that the 2-form $F$ is a NC field-strength tensor of some NC gauge theory. It
transforms covariantly under NC $U(1)$ gauge transformations
\begin{equation}
F' = U_\star\star F \star U_\star^{-1} .\nn 
\end{equation}
Here the finite NC gauge transformation is done with the matrix $U_\star$ and $U_\star$ can be a
function of the commutative gauge parameter $\alpha$. This fact and the explicit form of this
dependance is
of no relevance for the discussion of Hodge dual. Using (\ref{fStardx}) and (\ref{HodgeF}) one can
show that the transformation law of the Hodge-dual is given by
\begin{equation}
*_HF ' = U_\star \star (*_HF)\star U_\star^{-1}.
\end{equation}
That is, in the case of twist (\ref{AngTwist}) the natural definition of the NC Hodge dual form
transforms covariantly under the NC gauge transformations. This is an important result if one aims
to construct a gauge invariant action for the
NC gauge field, see the discussion in \cite{DJP}.  

The construction of Hodge dual in a curved space-time, in particular in the RN background
(\ref{dsRN}), we discuss in next section. 


\subsection{Angular twist in curved coordinates}

We have already mentioned that the vector field $X_2$ is nothing else but the
generator of rotations around the $z$-axis. Let us rewrite the twist (\ref{AngTwist}) in
the spherical coordinate system
\begin{eqnarray}
\mathcal{F} &=& e^{-\frac{i}{2}\theta ^{\alpha\beta}\partial_\alpha\otimes \partial_\beta} \nn\\
&=& e^{-\frac{ia}{2} (\partial_t\otimes\partial_\varphi - \partial_\varphi\otimes\partial_t)},
\label{AngTwist0Phi}
\end{eqnarray}
with $\alpha,\beta = t,\varphi$. Note that the twist has the same (\ref{AngTwist0Phi}) form in the
cylindrical coordinate system.

Now we rewrite all the formulas for the $\star$-product of functions and the differential
calculus in spherical coordinate system. Here we present the most important results, the rest can be
calculated easily:
\begin{eqnarray}
f\star g &=&  \mu \{ e^{\frac{ia}{2} (\partial_t\otimes \partial_\varphi - \partial_\varphi\otimes
\partial_t)}
f\otimes g \}\nn\\
&=& fg + \frac{ia}{2}
(\partial_t f(\partial_\varphi g) - \partial_t g(\partial_\varphi f)) + 
\mathcal{O}(a^2) ,\label{fStarg0Phi}\\
\d x^\mu \star f &=& f\star \d x^\mu = f \d x^\mu ,\label{fStardx0Phi}\\
\partial^\star _\mu f &=& \partial_\mu f .\label{partialStar0Phi}\\
\end{eqnarray}
Note that now $x^\mu = (t,r,\theta,\varphi)$.

In the case of the Hodge dual we have to use the definition of the Hodge dual in curved space-time.
This leads to
\begin{eqnarray}
*_HF = *_H(\frac{1}{2}F_{\mu\nu}\star \d x^\mu\wedge_\star
\d x^\nu) &=& \frac{1}{2}F_{\mu\nu}\star (\frac{1}{\sqrt{-g}}
\epsilon^{\mu\nu\rho\sigma}g_{\rho\alpha}g_{\sigma\beta}
\d x^\alpha\wedge_\star \d x^\beta) \nn\\
&=& \frac{1}{2\sqrt{-g}}\epsilon^{\mu\nu\rho\sigma}g_{\rho\alpha}g_{\sigma\beta} ( 
F_{\mu\nu}\star \d x^\alpha\wedge_\star \d x^\beta)\nn\\
&=& \frac{1}{2\sqrt{-g}}\epsilon^{\mu\nu\rho\sigma}g_{\rho\alpha}g_{\sigma\beta} 
F_{\mu\nu}\d x^\alpha\wedge \d x^\beta
, \label{HodgeF0Phi}
\end{eqnarray}
that is we obtained the commutative (undeformed) Hodge dual. In this calculation we used
(\ref{fStardx0Phi}). In addition, we used that the metric tensor $g_{\mu\nu}$ does
not
depend on $t,\varphi$ coordinates and therefore $g_{\mu\nu}\star f = g_{\mu\nu}\cdot f$ for an
arbitrary
function $f$. In a more general case, when the twist ${\cal F}$ is not a Killing twist for the
space-time metric $g_{\mu\nu}$, we cannot use
this definition of the Hodge dual, since in general $g_{\mu\nu}\star f \neq f\star g_{\mu\nu}$ and
$g_{\rho\alpha}\star g_{\sigma\beta}\neq g_{\sigma\beta}\star g_{\rho\alpha}$. These will spoil the
covariance of the $*_HF$ under the NC gauge transformations and make the construction of NC gauge
invariant action complicated, see \cite{PLGaugeGravity, DJP} for detailed discussion.
 
In the following we will work with the twist (\ref{AngTwist0Phi}) and we will develop the NC
scalar $U(1)_\star$ gauge theory on the RN background.


\section{Scalar $U(1)_\star$ gauge theory}

The twist (\ref{AngTwist0Phi}) enables us to study the behaviour of a NC scalar field in a
gravitational field of the Reissner–Nordstr\" om black hole.

Let us start from a more general action, describing the NC $U(1)_\star$ gauge theory of a complex
charged scalar field on an arbitrary background. The only requirement on the background is that
$\partial_t$ and $\partial_\varphi$ are Killing vectors.

If a one-form gauge field $\hat{A}=\hat{A}_\mu \star \d x^{\mu}$ is introduced to the model through
a minimal coupling, the relevant
action becomes
\begin{eqnarray}
S[\hat{\phi}, \hat{A}] &=& \int {\Big( \d {\hat{\phi}} - i\hat{A} \star {\hat{\phi}} \Big)}^+
\wedge_\star  *_H \Big( \d \hat{\phi} -i\hat{A} \star \hat{\phi} \Big) \nonumber \\
&& - \int \frac{\mu^2}{4!} \hat{\phi}^+\star \hat{\phi} \epsilon_{abcd}~ e^{a} \wedge_\star
e^b\wedge_\star e^c \wedge_\star e^d
- \frac{1}{4 q^2} \int (*_H \hat{F}) \wedge_\star \hat{F}. \label{NCActionGeometric}
\end{eqnarray}
The mass of the scalar field ${\hat{\phi}}$ is $\mu$, while its charge is $q$. The two-form
field-strength tensor is defined as
\begin{equation}
\hat{F} = \d \hat{A} - \hat{A} \wedge_\star
\hat{A} =\frac{1}{2} \hat{F}_{\m\nu}\star \d x^{\mu} \wedge_\star \d
x^{\nu}. \label{NCF} 
\end{equation}
In order to write the mass term for the scalar field $\hat{\phi}$ geometrically, we
introduced vierbein one-forms $e^a=e^a_\mu\star \d x^{\mu}$ and $g_{\mu\nu} = \eta_{ab}e_\mu^a\star
e_\nu^b$. In index notation, the action is of the form
\begin{eqnarray}
S[\hat{\phi}, \hat{A}] &=& S_\phi + S_A, \nn\\ 
S_\phi &=& \int \d ^4x \, \sqrt{-g}\star\Big( g^{\mu\nu}\star D_{\mu}\hat{\phi}^+ \star
D_{\nu}\hat{\phi} - \mu^2\hat{\phi}^+ \star\hat{\phi}\Big) , \label{SPhi}\\
S_A &=& -\frac{1}{4q^2} \int \d^4 x\,\sqrt{-g}\star g^{\alpha\beta}\star g^{\mu\nu}\star
\hat{F}_{\alpha\mu}\star \hat{F}_{\beta\nu} .\label{SA}
\end{eqnarray}
The scalar field $\hat{\phi}$ is a
complex charged
scalar field transforming in the fundamental representation of NC $U(1)_\star$. Its covariant
derivative is defined as
\begin{equation}
D_\mu\hat{\phi} = \partial_\mu\hat{\phi} - i \hat{A}_\mu\star \hat{\phi} \label{DPhi} . \nn
\end{equation}
The components of NC field-strength tensor follow from (\ref{NCF})
\begin{equation}
\hat{F}_{\mu\nu} = \partial_\mu \hat{A}_\nu - \partial_\nu \hat{A}_\mu -i [\hat{A}_\mu \ds \hat{A}_\nu]
.\label{F}
\end{equation}
The background gravitational field $g_{\mu\nu}$ is not specified for the moment. However, it is
important that its Killing vectors are $\partial_t$ and $\partial_\varphi$ since only in that case
the action (\ref{SA}) has this simple form. Note that $\star$-products in $\sqrt{-g}\star
g^{\alpha\beta}\star g^{\mu\nu}$ can all be removed since the twist (\ref{AngTwist0Phi}) does not
act on the metric tensor.

One can check that the actions (\ref{SPhi}) and (\ref{SA}) are invariant under the infinitesimal
$U(1)_\star$
gauge transformations\footnote{The actions (\ref{SPhi}) and (\ref{SA}) are also invariant under the
finite NC $U(1)_\star$ transformations defined as:
\begin{eqnarray}
\hat{\phi}' &=& U_\star\star \hat{\phi}, \nn\\
\hat{A}'_\mu &=& -U_\star \star\partial_\mu U_\star^{-1} + U_\star\star\hat{A}_\mu\star U_\star^{-1} ,\nn 
\end{eqnarray}
with $U_\star = e^{i\hat{\Lambda}}_\star = 1 + i\hat{\Lambda} +
\frac{1}{2}i\hat{\Lambda}\star i\hat{\Lambda} +\dots$.} defined in the following way:
\begin{eqnarray}
\delta_\star \hat{\phi} &=& i\hat{\Lambda} \star \hat{\phi}, \nn\\
\delta_\star \hat{A}_\mu &=& \partial_\mu\hat{\Lambda} + i[\hat{\Lambda} \ds \hat{A}_\mu],
\label{NCGaugeTransf}\\
\delta_\star \hat{F}_{\mu\nu} &=& i[\hat{\Lambda} \ds \hat{F}_{\mu\nu}],\nn\\
\delta_\star g_{\mu\nu} &=& 0.\nn
\end{eqnarray}
whith the NC gauge parameter $\hat{\Lambda}$.


\subsection{Seiberg-Witten map}

There are different approaches to construction of NC gauge theories. In this paper we use the
enveloping algebra approach \cite{jssw} and the Seiberg-Witten (SW) map \cite{seiberg}. SW map
enables to express NC variables as functions of the coresponding commutative variables. In this
way, the problem of charge quantization in $U(1)_\star$ gauge theory does not exist. In the case of
NC Yang-Mills gauge theories, SW map guarantees that the number of degrees of freedom in the NC
theory is the same as in the corresponding commutative theory. That is, no new degrees of freedom
are introduced.

Using the SW-map NC fields can be expressed as function of corresponding commutative fields
and can be expanded in orders of the deformation parameter $a$. Expansions for an arbitrary Abelian
twist deformation are known to all orders \cite{PLSWGeneral}. Applying these results to the twist
(\ref{AngTwist0Phi}), 
expansions of fields up to first
order in the deformation parameter $a$ follow. They are given by:
\begin{eqnarray}
\hat{\phi} &=& \phi -\frac{1}{4}\theta^{\rho\sigma}A_\rho(\partial_\sigma\phi + D_\sigma
\phi), \label{HatPhi}\\
\hat{A}_\mu &=& A_\mu -\frac{1}{2}\theta^{\rho\sigma}A_\rho(\partial_\sigma A_{\mu} +
F_{\sigma\mu}), \label{HatA}\\
\hat{F}_{\mu\nu} &=& F_{\mu\nu} - \frac{1}{2}\theta^{\rho\sigma}A_{\rho}(\partial_\sigma F_{\mu\nu}
+ D_\sigma F_{\mu\nu})+\theta^{\rho\sigma}F_{\rho\mu}F_{\sigma\nu}. \label{HatFmunu}
\end{eqnarray}
The $U(1)$ covariant derivative of $\phi$ is defined as $D_\mu \phi = (\partial_\mu - i A_\mu)
\phi$, while $D_\sigma F_{\mu\nu} = \partial_\sigma F_{\mu\nu}$ in the case of $U(1)$ gauge
theory. The commutative complex scalar field $\phi$
can be decomposed into two real scalar fields $\phi_1,\phi_2$ in the usual way
$\phi = \frac{1}{\sqrt{2}}(\phi_1 + i\phi_2)$. It is important to note that the coupling constant
$q$
between fields $\phi$ and $A_\mu$, the charge of $\phi$, is included into $A_\mu$, namely $A_\mu =
qA_\mu$, compare also (\ref{SA}).


\subsection{Expanded actions and equtions of motion}

Using the SW-map solutions and expanding the $\star$-products in (\ref{SPhi}) and (\ref{SA}) we
find the action up to first order in the deformation parameter $a$. It is given by
\begin{eqnarray}
S &=& \int
\d^4x\sqrt{-g}\,
\Big( -\frac{1}{4q^2}g^{\mu\rho}g^{\nu\sigma}F_{\mu\nu}F_{\rho\sigma}
+ g^{\mu\nu}D_\mu\phi^+D_\nu\phi -\mu^2\phi^+\phi \nonumber\\
&&
+\frac{1}{8q^2}g^{\mu\rho}g^{\nu\sigma}\theta^{\alpha\beta}(F_{\alpha\beta}F_{\mu\nu}F_{\rho\sigma}
-4F_{\mu\alpha}F_{\nu\beta}F_{\rho\sigma}) +\frac{\mu^2}{2}\theta^{\alpha\beta}F_{\alpha\beta}\phi^+\phi \label{SExp}\\
&& + \frac{\theta^{\alpha\beta}}{2}g^{\mu\nu}\big( -\frac{1}{2}D_\mu\phi^+F_{\alpha\beta}
D_\nu\phi
+(D_\mu\phi^+)F_{\alpha\nu}D_\beta\phi + (D_\beta\phi^+)F_{\alpha\mu}D_\nu\phi\big) \Big)
.\nn 
\end{eqnarray}

Now we vary the action (\ref{SExp}) to calculate the equations of motion. Varying the action
with respect to $\phi^+$ we obtain
\begin{equation}
\begin{split} 
&  g^{\mu \nu} \bigg( (\partial_{\mu} - iA_{\mu})D_{\nu} \phi - \Gamma^{\lambda}_{\mu
\nu}D_{\lambda} \phi \bigg) -\mu^2\phi \\
& +\frac{\mu^2}{2}\theta^{\alpha\beta}F_{\alpha\beta}\phi -\frac{1}{4} \theta^{\alpha \beta} g^{\mu \nu} \bigg(  (\partial_{\mu} - iA_{\mu})(
F_{\alpha \beta}D_{\nu} \phi)
- \Gamma^{\lambda}_{\mu \nu} F_{\alpha \beta}  D_{\lambda} \phi   \\
& - 2  (\partial_{\mu} - iA_{\mu})( F_{\alpha \nu}D_{\beta} \phi)  + 2
\Gamma^{\lambda}_{\mu \nu} F_{\alpha \lambda}  D_{\beta} \phi  -2  (\partial_{\beta} -
iA_{\beta})( F_{\alpha \mu}D_{\nu} \phi) \bigg)  = 0.  \label{EoMPhi}
\end{split}
\end{equation}
The $U(1)$ covariant derivative $D_\mu\phi$ has been defined in the previous subsection. Since the
backround metric $g_{\mu\nu}$ is not flat, the corresponding Christoffel symbols
$\Gamma^{\lambda}_{\mu \nu}$ appear in the equation of motion.

Varying the action (\ref{SExp}) with respect to $A_\lambda$ we obtain
\begin{eqnarray}
&& \partial_\mu F^{\mu\lambda} + \Gamma^{\rho}_{\mu \rho} F^{\mu\lambda}
+\theta^{\alpha\beta}
\Big( -\frac{1}{2}\big( \partial_\mu (F_{\alpha\beta}F^{\mu\lambda}) + \Gamma^{\rho}_{\mu \rho}
F_{\alpha\beta}F^{\mu\lambda} \big) \nn\\
&& + \partial_\mu (F_{\alpha}^{\ \mu}F_\beta^{\ \lambda}) +
\Gamma^{\rho}_{\mu \rho}F_{\alpha}^{\ \mu}F_\beta^{\ \lambda}
-\partial_\alpha(F_{\beta\mu}F^{\mu\lambda}) \Big) \nn\\
&& +\theta^{\alpha\lambda}
\Big( \frac{1}{2}\big( \partial_\mu (F_{\beta\nu}F^{\mu\nu})
+\Gamma^\rho_{\mu\rho}F_{\beta\nu}F^{\mu\nu}\big) -\frac{1}{4}\partial_\alpha(F_{\mu\nu}F^{\mu\nu})
\Big) =
q^2j^\lambda ,\label{EoMA}
\end{eqnarray}
with
\begin{eqnarray}
j^\lambda &=& (1-\frac{1}{4}\theta^{\alpha\beta}F_{\alpha\beta})j^\lambda_{(0)}
-\frac{1}{2}\theta^{\alpha\beta}g^{\lambda\mu} \Big( -F_{\alpha\mu}j^{(0)}_\beta
-\partial_\alpha(D_\mu\phi^+D_\beta\phi
+D_\beta\phi^+D_\mu\phi)\Big)\nn\\
&& -\frac{1}{2}\theta^{\alpha\lambda}\Big( -F_{\alpha\nu}j^{\nu(0)}
+g^{\mu\nu}\partial_\alpha(D_\mu\phi^+D_\nu\phi)\label{j}\\
&& -\partial_\nu\big( g^{\mu\nu}(D_\mu\phi^+D_\alpha\phi
+D_\alpha\phi^+D_\mu\phi)\big) -\Gamma_{\nu\rho}^\rho\big( g^{\mu\nu}(D_\mu\phi^+D_\alpha\phi
+D_\alpha\phi^+D_\mu\phi)\big) \Big).\nn
\end{eqnarray}
The zeroth order current is defined as
\begin{equation}
j^\lambda_{(0)} = ig^{\mu\lambda}\big( D_\mu\phi^+ \phi -\phi^+D_\mu\phi \big) ,\label{j0} 
\end{equation}
and $j^{(0)}_\beta = g_{\beta\lambda}j^\lambda_{(0)}$. Once again, the Christoffel symbols
corresponding to the metric $g_{\mu\nu}$ appear in the equation.


\section{Scalar field in the Reissner–Nordstr\" om background}

Finally, let us specify the gravitational background to be that of a charged non-rotating black
hole, the Reissner–Nordstr\" om (RN) black hole. Since we are interested in the QNMs of the scalar
field, we will only consider the equation (\ref{EoMPhi}) and assume that the gravitational
field $g_{\mu\nu}$ and the $U(1)$ gauge field $A_\mu$ are fixed to be the gravitational field and
the
electromagnetic field of the RN black hole.

We write the RN metric tensor once again
\begin{equation}
g_{\mu\nu}=
\begin{bmatrix}
f & 0 & 0 & 0  \\
0 & -\frac{1}{f} & 0 & 0 \\
0 & 0 & -r^2 & 0  \\
0 & 0 & 0 & -r^2\sin^2\theta
\end{bmatrix} \label{RNmetric}
\end{equation}
with $f = 1-\frac{2MG}{r}+\frac{Q^2G}{r^2}$ and $M$ is the mass of the RN black hole,
while $Q$ is the charge of the RN black hole. The corresponding non-zero Christoffel symbols are:
\begin{eqnarray}
&& \Gamma^r_{tt} = \frac{f}{r}(\frac{MG}{r} - \frac{Q^2G}{r^2}),\quad \Gamma^r_{rr} =
-\frac{\frac{MG}{r} - \frac{Q^2G}{r^2}}{rf} = -\Gamma^t_{rt}, \nn\\
&& \Gamma^r_{\varphi\varphi} = -rf\sin^2\theta,\quad 
\Gamma^r_{\theta\theta} = -rf,\nn \\
&& \Gamma^\theta_{\varphi\varphi} = -\sin\theta\cos\theta,\quad
\Gamma^\varphi_{\theta\varphi} = \cot\theta, \quad \Gamma^\theta_{r\theta} = \Gamma^\varphi_{\varphi r}
=\frac{1}{r} .\label{ChrSymb}
\end{eqnarray}
The RN black hole is non-rotating, therefore the only non-zero component of the gauge field is the
scalar potential
\begin{equation}
A_0 = -\frac{qQ}{r}. \label{A0}
\end{equation}
The corresponding electric field is given by
\begin{equation}
F_{r0} = \frac{qQ}{r^2}. \label{Fr0}
\end{equation}
The only non-zero components of the NC deformation parameter $\theta^{\alpha\beta}$ are
$\theta^{t\varphi}=
-\theta^{\varphi t}=a$. Inserting all these into (\ref{EoMPhi}) gives the following equation
\begin{eqnarray}
&&
\Big( \frac{1}{f}\partial^2_t -\Delta + (1-f)\partial_r^2 
+\frac{2MG}{r^2}\partial_r + 2iqQ\frac{1}{rf}\partial_t -\frac{q^2Q^2}{r^2f} -\mu^2 \Big)\phi
\nonumber\\
&& +\frac{aqQ}{r^3}
\Big( (\frac{MG}{r}-\frac{GQ^2}{r^2})\partial_\varphi
+ rf\partial_r\partial_\varphi \Big) \phi =0 .\label{EomPhiExp1}
\end{eqnarray}
Note that $\Delta$ is the usual Laplace operator. 

In order to solve this equation we assume an ansatz
\begin{equation}
\phi_{lm}(t,r,\theta,\varphi) = R_{lm}(r)e^{-i\omega t}Y_l^m(\theta, \varphi) \label{AnsatzPhi} 
\end{equation}
with spherical harmonics $Y_l^m(\theta, \varphi)$. Inserting (\ref{AnsatzPhi}) into
(\ref{EomPhiExp1}) leads to an equation for the radial function $R_{lm}(r)$
\begin{eqnarray}
&& f R_{lm}'' + \frac{2}{r}\big( 1-\frac{MG}{r}\big) R_{lm}' - \Big( \frac{l(l+1)}{r^2} 
- \frac{1}{f}(\omega - \frac{qQ}{r})^2  + \mu^2 \Big)R_{lm} \nn\\
&& -ima\frac{qQ}{r^3}\Big( (\frac{MG}{r}-\frac{GQ^2}{r^2})R_{lm} + rf R_{lm}' \Big) =0
.\label{EoMR} 
\end{eqnarray}
The zeroth oder of this equation corresponds to the equation for the radial function $R_{lm}$ in
\cite{Hod:2010hw, QNMode}. 


\section{Solutions for QNMs}

We are interested in a special solution of equation (\ref{EoMR}), namely the quasinormal mode
solution. This solution describes damped oscillations of a perturbed black hole, as the black hole
goes through the ringdown phase. A set of the boudary condition which leads to this solution is the
following: at the horizon, the QNMs are purely incoming, while in the infinity the QNMs are purely
outgoing. In order to find the spectrum of QNMs in our model, let us firstly rewrite equation
(\ref{EoMR}) in a more convenient way.

Taking into account that the outer and inner horizon of RN black hole are given by $r_{\pm} = 
GM\pm\sqrt{G^2 M^2 - G Q^2}$, we introduce the following variables and abbreviations
\begin{eqnarray}
&& x= \frac{r-r_+}{r_+}, \qquad \tau = \frac{r_+ -r_-}{r_+} ,\lab{Abr1}\\
&& k= 2 \omega r_+ - qQ, \qquad  \kappa = \frac{\omega k - \mu^2 r_+}{\sqrt{\omega^2 -
\mu^2}}\label{Abr2}\\
&& \Omega = \frac{\omega r_+ - qQ}{\sqrt{G^2 M^2 - G Q^2}} r_+ = 2
\frac{\omega r_+ - qQ}{\tau}. \label{Abr3}
\end{eqnarray} 
Then the radial equation of motion (\ref{EoMR}) reduces to 
\begin{equation} 
\begin{split}  \label{findsol1}
& x(x+\tau) \frac{{\d}^2 R}{\d x^2} +\bigg( 2x +\tau - ima \frac{qQ}{r_+}
\frac{x(x+\tau)}{{(x+1)}^2} \bigg) \frac{{\d} R}{\d x}   
-\bigg[  l(l+1) + \mu^2 r_+^2 {(x+1)}^2   \\
& -  \frac{{{(\omega r_+^2 x^2 + k r_+ x + r_+ \frac{\Omega \tau}{2})}^2}}{r_+^2 x(x+\tau)}  
-ima \frac{GMqQ}{r_+^2 {(x+1)}^2} + 
ima \frac{GqQ^3}{r_+^3 {(x+1)}^3}   \bigg]R =0 .
\end{split}
\end{equation}
In order to simplify writing, instead of $R_{lm}$ we just write $R$ for the radial part of the
scalar field, keeping in mind that it depends on $l$ and $m$. 

The equation (\ref{findsol1}) can be
treated analytically in the near extremal case $\tau \ll 1$ or
$\frac{Q^2}{GM^2}\to 1$.
  
The strategy that we adopt here is similar to that in \cite{Hod:2010hw}. It allows us to analyse the equation (\ref{findsol1}) in two different
regions, one being relatively far from the horizon, $x\gg \tau$, and the other  being relatively
close to the
horizon, $x\ll 1$. It is important to emphasize that these two regions have to be
chosen in such a way as to
ensure that  their  region of a common overlap  exists (or is likely being close to exist).
The restriction to a near extremal limit that we make, as well as  an appropriate  choice of a range of the system
parameters makes this possible.

With a presence of a common overlapping  region being guaranteed, we  extrapolate the solutions obtained in two separate
regions all the way up (or down) to this   region of a common overlap. The subsequent comparison of the extrapolated solutions is then used to fix  the unknown constants of integration.

In the region $x \gg \tau$ equation (\ref{findsol1}) can be approximated by 
\begin{equation} 
x^2 \frac{{\d}^2 R}{\d x^2} + 2x \frac{{\d} R}{\d x} -\bigg[ l(l+1) + \mu^2 r_+^2 {(x+1)}^2
-\omega^2 r_+^2 x^2 -2\omega  r_+ k x -k^2 \bigg]R =0. \label{Reg1}
\end{equation}
In our analysis we set the scale of the NC parameter $a$ to  be of the order of the
"extremality" $\tau$. In
that case, all NC corrections in equation (\ref{Reg1}) vanish.  This is 
 due to the approximation being used, $a \sim \tau,$ and due to the region
being considered, $x \gg \tau.$  Namely,  being proportional to $a$ and having powers of $x$ in the denominator, these terms  give rise to the least dominant  contributions in the region  $x  \gg \tau $.  That is, they
are higher order corrections in  $\frac{\tau}{x}$.

Next we consider the region  $x \ll 1$. Picking up all relevant terms and introducing a
new conveniently chosen variable $y = -\frac{x}{\tau}$, from (\ref{findsol1}) we obtain
\begin{equation}
\label{Reg2}
\begin{split}
& y(1-y) \frac{{\d}^2 R}{\d y^2} + (1-2y) \frac{{\d} R}{\d y}   
+\bigg[  l(l+1) + \mu^2 r_+^2   \\
& +\frac{ {(k  y - \frac{\Omega}{2})}^2 +\omega r_+ \tau (\Omega y^2 -2k y^3)}{y(1-y)}  - ima \frac{GMqQ}{r_+^2} + ima
\frac{GqQ^3}{r_+^3} \bigg]R =0.
\end{split}	
\end{equation}
We can immediately see that NC corrections survive in this region. In this way, our approximation
results in NC corrections near the horizon where the gravitational field is strong. Far from the
horizon, where the gravitational field is weak, the NC corrections are not present. These results are
consistent with the assumption that in the strong gravitational field the space-time structure
becomes deformed.

We point out that the approximation  $ a\sim\tau $ was essential in obtaining the equations (\ref{Reg1}) and (\ref{Reg2}).  
Besides this one, an additional approximation has been used to get (\ref{Reg1}) and (\ref{Reg2}). This latter approximation is dictated by the character of the region  in which the analysis has been made ($x  \gg \tau $ or $x\ll 1$). However, note that  since $\tau$ can be arbitrarily small,   it does not severely constrain  the value of the NC parameter $a$.

In the first region, $x\gg \tau$, the substitution $R(x) = e^{-\alpha x} x^{\beta} g(x)$
transforms equation (\ref{Reg1}) into a  confluent hypergeometric equation with the solution
\begin{equation}
\label{Reg1a}
\begin{split}
& R(x) = C_1 {(2 \alpha)}^{\frac{1}{2} + i\sigma} x^{-\frac{1}{2} + i\sigma}~ e^{- \alpha x} 
M \bigg( \frac{1}{2} + i\sigma + i\kappa, 1+ 2i\sigma, 2 \alpha x \bigg) \\
& + C_2 {(2 \alpha)}^{\frac{1}{2} - i\sigma} x^{-\frac{1}{2} - i\sigma}~ e^{-\alpha x} 
M \bigg(\frac{1}{2} - i\sigma + i\kappa, 1- 2i\sigma, 2 \alpha x \bigg),
\end{split}	
\end{equation}
where $\alpha = i \sqrt{\omega^2 - \mu^2} ~r_+$ and $\sigma = \sqrt{k^2 
-\mu^2 r_+^2 - {(l+\frac{1}{2})}^2}$ and  $M$ is the Kummer's hypergeometric function. 
As a matter of fact, in order to get a  confluent hypergeometric equation, 
the parameters $\alpha$ and $\beta$  can actually take  two values each, $\alpha = \pm  i \sqrt{\omega^2 - \mu^2} ~r_+ $ and 
$\beta = -\frac{1}{2} \pm i\sigma$, respectively. However, as far as the physical arguments go, not each of these values is acceptable. The important point to realise here is that the solutions obtained in two regions, $x \gg \tau$ and $x \ll 1$, not only need to satisfy the proper boundary conditions, but they also need to match mutually
in the region of  their common  overlap. This means that,  when extrapolated, they have to converge to  expressions that have the same analytical form and the proper choice of the parameter $ \beta$, as well as the parameters $\lambda_1$ and $\lambda_2$ (see below) appears to be crucial   to this purpose. For these reasons, all these  parameters have to be chosen carefully. In particular,
 $\beta$ is chosen as $\beta = -\frac{1}{2} +i\sigma $.
If instead the parameters are not selected carefully, two solutions obtained by extrapolation will not converge to the expressions of the same kind and thus two solutions will not be able to match in their common  region of overlap.
As for $\alpha,$ it is the QNMs  boundary condition at far infinity which determines the choice of its sign.

In the second region, $x\ll 1$, the substitution $R(y) = y^{\lambda_1} {(1-y)}^{\lambda_2} F(y)$
transforms equation (\ref{Reg2}) into a hypergeometric equation with the solution
\begin{equation}  
R(y) = y^{-i\frac{\Omega}{2}} {(1-y)}^{i (\frac{\Omega}{2} -k)}~ 
F \bigg( \frac{1}{2} + i\sigma - ik + \tilde{\rho}, 
\frac{1}{2} - i\sigma - ik - \tilde{\rho}, 1- i\Omega; ~y \bigg) \label{Reg2a}
\end{equation}
where $\tilde{\rho} = \frac{am}{2\sigma}(\frac{G qQ^3}{r_+^3} -\frac{GMqQ}{r_+^2})  +i \frac{\sqrt{G^2 M^2 - G Q^2}  \omega}{2 \sigma } (k + \frac{\Omega}{2}) ~$ and
$\sigma = \sqrt{k^2 - \mu^2 r_+^2 -(l+ \frac{1}{2})^2}$. Note here that $~\tau = \sqrt{G^2 M^2 - G Q^2}/r_+.$

As in the previous situation,   the  parameters $\lambda_1$ and $\lambda_2$ 
 for which the equation (\ref{Reg2}) reduces to a hypergeometric equation, namely  $\lambda_1 = \pm i \frac{\Omega}{2}$ 
 and $\lambda_2 =i(\frac{\Omega}{2} \pm k),$ are not unique. Instead,  they may acquire two possible values  each and, as already mentioned before, need to be selected carefully in order to meet the  physical demands.
The value of $\lambda_1$ is chosen with the minus sign, $\lambda_1 = - i \frac{\Omega}{2}$, 
so to comply with the QNMs  boundary condition at the horizon.
Furthermore, the parameter  $\lambda_2$  has been selected as  $\lambda_2 =i(\frac{\Omega}{2} -k)$. 
With this choice, the extrapolated solutions obtained in two regions, $x \gg \tau$ and $x \ll 1,$
can be  matched together.

Once $\lambda_1$ has been selected, the general solution of equation (\ref{Reg2}) may be written as a linear combination of two independent solutions of the hypergeometric equation.
The requirement  of having a purely ingoing wave at the horizon forces us to drop out one of
these two independent functions, resulting in a solution of the form  (\ref{Reg2a}).

Extrapolation of both solutions to the mutual overlap region $\tau \ll x \ll 1$ leads to
\begin{equation}  
R(x)_{\mbox{\tiny Reg1}}  ~\simeq  ~ C_1 {(2 i \sqrt{\omega^2 - \mu^2}
~r_+)}^{\frac{1}{2} +
i\sigma} x^{-\frac{1}{2} + i\sigma} 
+ C_2 {(2 i \sqrt{\omega^2 - \mu^2} ~r_+)}^{\frac{1}{2} - i\sigma} x^{-\frac{1}{2} -
i\sigma}\label{Reg1b}
\end{equation}
and
\begin{eqnarray}  
R(x)_{\mbox{\tiny Reg2}} &\simeq &  {(-1)}^{-i\frac{\Omega}{2}} \Bigg[ \frac{\Gamma(1-i\Omega) \Gamma(-2i\sigma - 2
\tilde{\rho})
\tau^{\frac{1}{2} + i\sigma  + \tilde{\rho}}}{\Gamma(\frac{1}{2}-i\sigma -
ik-\tilde{\rho})
\Gamma(\frac{1}{2}-i\sigma + ik -  i\Omega -\tilde{\rho})} x^{-\frac{1}{2} - i\sigma}  \nonumber \\
&&+  \frac{\Gamma(1-i\Omega) \Gamma(2i\sigma + 2 \tilde{\rho}) \tau^{\frac{1}{2} 
- i\sigma  - \tilde{\rho}}}{\Gamma(\frac{1}{2} + i\sigma - ik +  \tilde{\rho})
\Gamma(\frac{1}{2}+i\sigma + ik -  i\Omega +\tilde{\rho})}  x^{-\frac{1}{2} + i\sigma} \Bigg]
.\label{Reg2b}
\end{eqnarray}
By matching these two expressions, it is possible to fix the so far unknown integration constants
$C_1$ and $C_2$. We obtain
\begin{equation}
C_1 = {(-1)}^{-i\frac{\Omega}{2}} \frac{\Gamma(1-i\Omega) \Gamma(2i\sigma + 2 \tilde{\rho}) \tau^{\frac{1}{2} - i\sigma - \tilde{\rho}}}{\Gamma(\frac{1}{2}+i\sigma - ik +  \tilde{\rho})
\Gamma(\frac{1}{2}+i\sigma + ik -  i\Omega +\tilde{\rho})} {(2i \sqrt{\omega^2 - \mu^2}
r_+)}^{-\frac{1}{2} - i\sigma} \label{C1}
\end{equation}
and
\begin{equation}
C_2 = {(-1)}^{-i\frac{\Omega}{2}} \frac{\Gamma(1-i\Omega) \Gamma(-2i\sigma - 2 \tilde{\rho}) \tau^{\frac{1}{2} + i\sigma
+ \tilde{\rho}}}{\Gamma(\frac{1}{2}-i\sigma - ik -  \tilde{\rho})
\Gamma(\frac{1}{2}-i\sigma + ik -  i\Omega -\tilde{\rho})} 
{(2i \sqrt{\omega^2 - \mu^2} r_+)}^{-\frac{1}{2} + i\sigma} .\label{C2}
\end{equation}
Note that in obtaining the  expression (\ref{Reg2b}) we have used the linear transformation 
\begin{eqnarray}  \label{lintransf1}
  F(a,b,c;y)  & = & \frac{\Gamma(c) \Gamma(b-a)}{\Gamma(b) \Gamma(c-a)} {(-y)}^{-a} F(a,1-c+a,1-b+a; \frac{1}{y})  \nonumber \\
     &  +  &  \frac{\Gamma(c) \Gamma(a-b)}{\Gamma(a) \Gamma(c-b)} {(-y)}^{-b}
      F(b, 1-c+b, 1-a+b; \frac{1}{y}).
\end{eqnarray}
Taking the limit $x \gg \tau$ makes $1/y$ 
approach  zero, and one can use the standard Taylor expansion of the function $F(a,b,c; y)$
around zero to get the most dominant contributions  coming from the second region. 

Consider again the solution (\ref{Reg1a}) in the region $x \gg \tau$. From the asymptotic
behaviour of the confluent hypergeometric functions 
\begin{equation}
M(a, b, z) ~\xrightarrow{|z| \rightarrow \infty }
~\Gamma(b) \bigg(  \frac{e^z z^{a-b}}{\Gamma(a)}
+ \frac{{(-z)}^{-a}}{\Gamma(b-a)} \bigg) + O({|z|}^{a-b-1}) \label{AsimptoticM}
\end{equation}
the asymptotic behavior of the solution (\ref{Reg1a}) follows
\begin{equation}
\label{Reg1Asympt}
\begin{split}
R(x) &\sim \bigg[C_1 {(2 \alpha)}^{ i\kappa}  \frac{\Gamma(1+2i\sigma)}{\Gamma(\frac{1}{2} +i\sigma
+i \kappa)} x^{-1 +i\kappa}
+ C_2 {(2 \alpha)}^{ i\kappa} \frac{\Gamma(1-2i\sigma)}{\Gamma(\frac{1}{2} -i\sigma +i \kappa)}
x^{-1  +i\kappa} \bigg]~ e^{ \alpha x}  \\
&+ \bigg[{(-1)}^{-\frac{1}{2} - i\sigma -i\kappa}C_1 {(2 \alpha)}^{ -i\kappa} 
\frac{\Gamma(1+2i\sigma)}{\Gamma(\frac{1}{2} +i\sigma -i \kappa)} x^{-1 -i\kappa} \\
& + {(-1)}^{-\frac{1}{2} + i\sigma -i\kappa}C_2 {(2 \alpha)}^{ -i\kappa}
\frac{\Gamma(1-2i\sigma)}{\Gamma(\frac{1}{2} -i\sigma -i \kappa)} x^{-1  -i\kappa} \bigg]~
e^{-\alpha x} + O(x^{-1}) .
\end{split}
\end{equation}
Recalling the QNMs boundary condition of having a purely outgoing wave at far infinity, it is clear that the second square bracket
in the expression (\ref{Reg1Asympt}) has to vanish.
The equation (\ref{Reg1Asympt}) thus  provides another constraint on the constants $C_1$ and $C_2$, namely 
\begin{equation}  
C_1  \frac{\Gamma(1+2i\sigma)}{\Gamma(\frac{1}{2} +i\sigma -i \kappa)} {(-1)}^{ - i\sigma } 
+ C_2  \frac{\Gamma(1-2i\sigma)}{\Gamma(\frac{1}{2} -i\sigma -i \kappa)} {(-1)}^{  i\sigma }  =0
.\label{C1C2constraint}
\end{equation}
Combining this constraint with (\ref{C1}) and (\ref{C2}), the  quantization condition emerges
that determines the black hole QNMs spectrum. It is given by
\begin{equation}
\label{QNMcondition}
\begin{split}
&  \frac{\Gamma(1-2i\sigma) \Gamma(-2i\sigma - 2 \tilde{\rho}) }{\Gamma(\frac{1}{2} -i\sigma -ik -
\tilde{\rho} ) \Gamma(\frac{1}{2} -i\sigma +ik - i \Omega -\tilde{\rho} ) \Gamma(\frac{1}{2}
-i\sigma -i \kappa)}   \\
& = - \frac{\Gamma(1+2i\sigma) \Gamma(2i\sigma + 2 \tilde{\rho}) \tau^{-2
\tilde{\rho}}}{\Gamma(\frac{1}{2} +i\sigma -ik + \tilde{\rho} ) \Gamma(\frac{1}{2} +i\sigma +ik - i
\Omega +\tilde{\rho} ) \Gamma(\frac{1}{2} +i\sigma -i \kappa)} \\
& \times {\Big( -2 i \sqrt{\omega^2 -
\mu^2}
~r_+ \tau \Big) }^{ -2i\sigma} .
\end{split}
\end{equation}
In general, this condition cannot be solved analytically. In the next section we discuss a special
choice of parameters where the analytic solution is possible. In the following, we present
some numerical results for the QNMs frequences, obtained by Wolfram Mathematica and resulting from
the analytic condition 
(\ref{QNMcondition}). In particular, we shall be concerned with the fundamental quasinormal mode.
However, before we  proceed to discuss the properties of QNMs frequences and describe the behaviour of the fundamental mode in terms of certain system parameters, like the mass and charge of the scalar probe, we explain  why the fundamental tone is so important.

To begin with, the fundamental mode in the QNMs
spectrum has the frequency with the lowest absolute value of the imaginary part (with  a sign of the
imaginary part of course being such that it guarantees stability of a black hole). Since the
imaginary part of the frequency is proportional to a decay rate of a perturbation, it is clear that 
the modes with larger  imaginary part of the frequency will more quickly die out, and the
fundamental
mode will be the one that will dominate the perturbation signal.
Therefore, in the forthcoming experiments that are going to utilise the current  gravitational antennas 
(such as LIGO, VIRGO, LISA) for observing   gravitational signals from black holes, the most
dominant contribution to these signals will come from the fundamental mode of the QNMs spectrum. 

 From the reasons just explained,  we give  separately  the dependence
of the fundamental QNMs frequency $\omega$ (i.e. its real and imaginary part) on the charge of the scalar field $q$
and on the mass of the scalar field $\mu$. In order to make comparison, we plot the results for two
choices
of "extremality": $\frac{Q}{M}=0.9999$, corresponding to the "extremality" parameter $\tau
=0.0278891617$ and $\frac{Q}{M}=0.999999$, corresponging to $\tau=0.0028244321$. For simplicity we set $G=1$. When deriving (\ref{QNMcondition}) we used the approximation that the NC deformation parameter $a$ is of the same order as the extremality $\tau$. Therefore, in the case $\frac{Q}{M}=0.9999$ we set $a=0.1$ while $a=0.01$ corresponds to the case $\frac{Q}{M}=0.999999$.
The $q$ dependance of $ \mbox{Re}\,\omega$ and
$\mbox{Im}\,\omega$ is presented in Figures 1 and 2. In this case we assumed that the mass of the scalar
field is fixed at $\mu=0.05$. 

\begin{center}
\begin{tabular}{lll}
\includegraphics[scale=0.3]{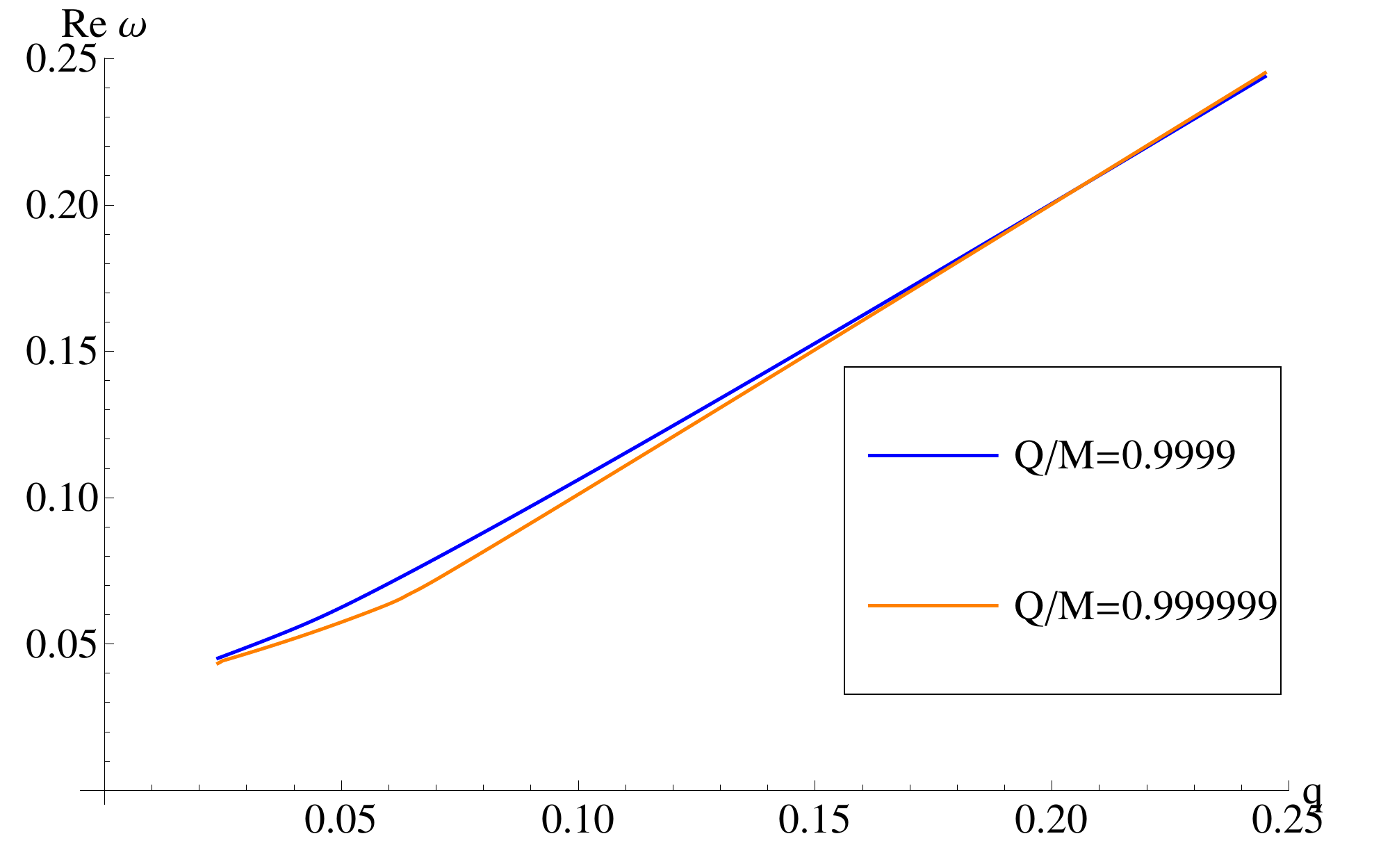}& & \includegraphics[scale=0.3]{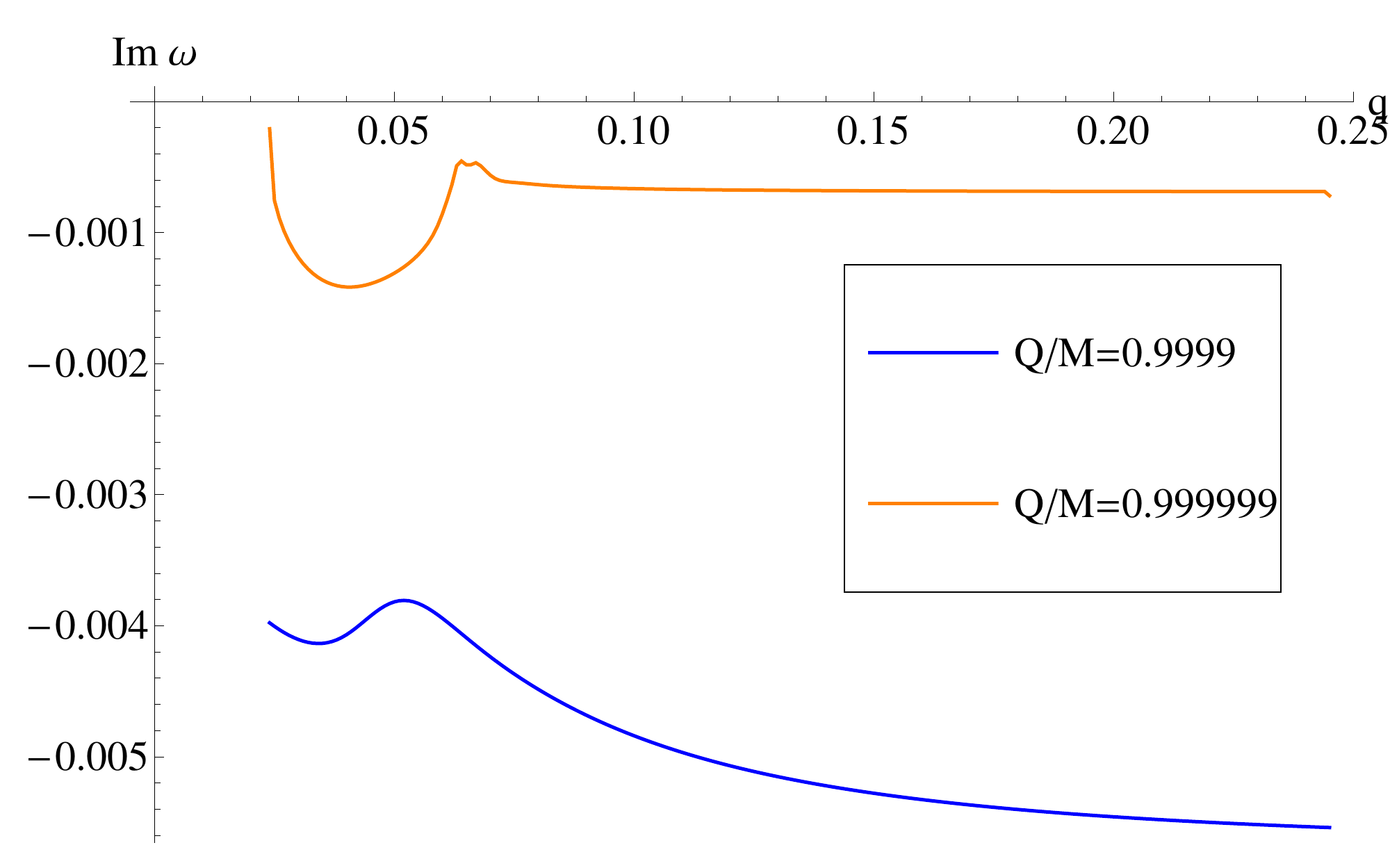}\\
{\scriptsize Figure 1: Dependance of $ \mbox{Re}\,\omega$ on the charge $q$ }& & {\scriptsize Figure 2: Dependance of $ \mbox{Im}\,\omega$ on the charge $q$}\\
\hspace*{1.3cm}{\scriptsize of the scalar field with the mass } & & \hspace*{1.2cm} {\scriptsize of the scalar field with the mass}\\
\hspace*{1.3cm}{\scriptsize $\mu=0.05$, $l=1$.} & & \hspace*{1.3cm}{\scriptsize $\mu=0.05$, $l=1$.}
\end{tabular} 
\end{center}

Likewise, the dependance of the fundamental QNMs frequency $\omega$ on the mass of the scalar
field $\mu$, for the
fixed charge  $q=0.075$,  is shown in Figures 3 and 4.
\begin{center}
\begin{tabular}{lll}
\includegraphics[scale=0.3]{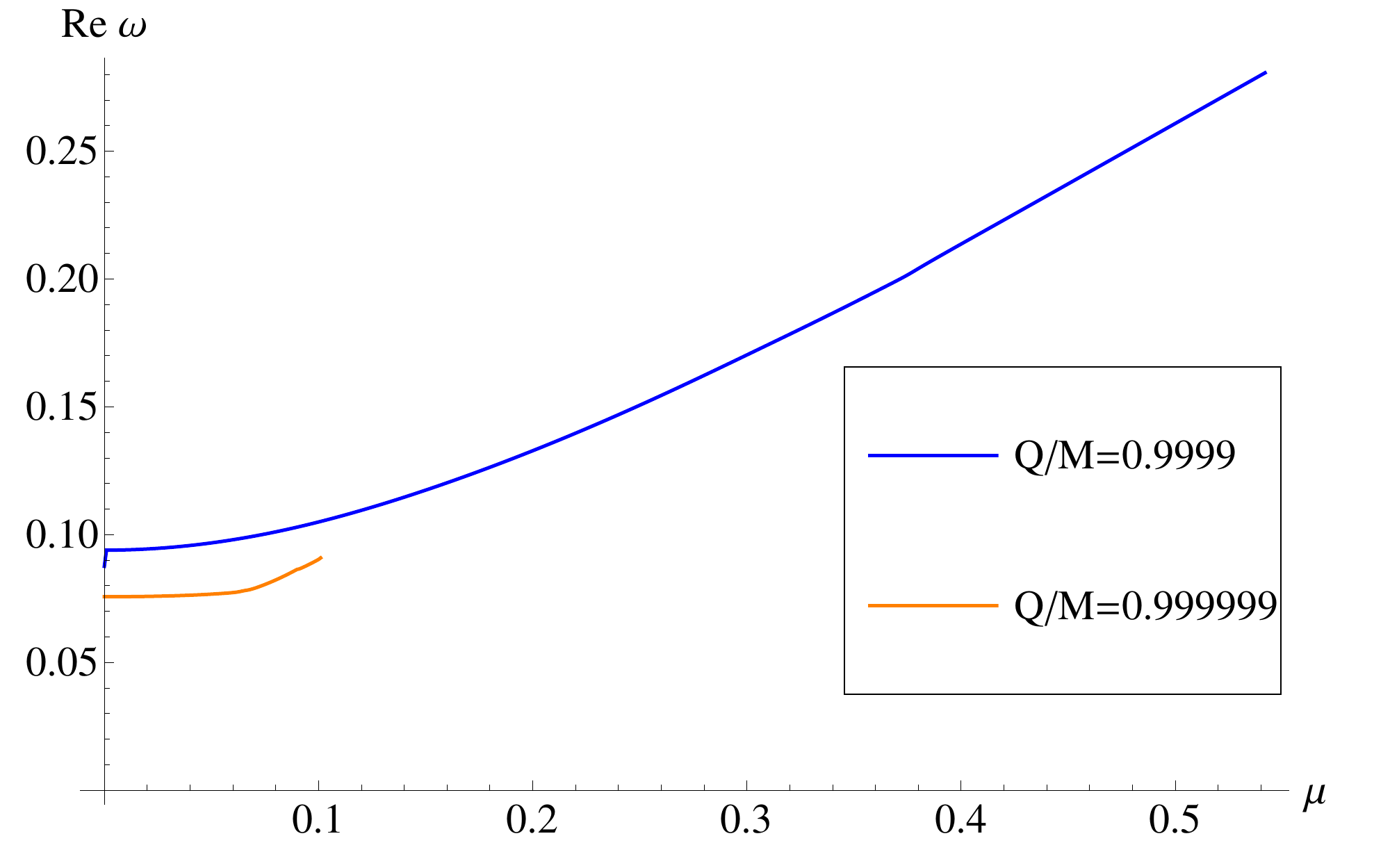}& & \includegraphics[scale=0.3]{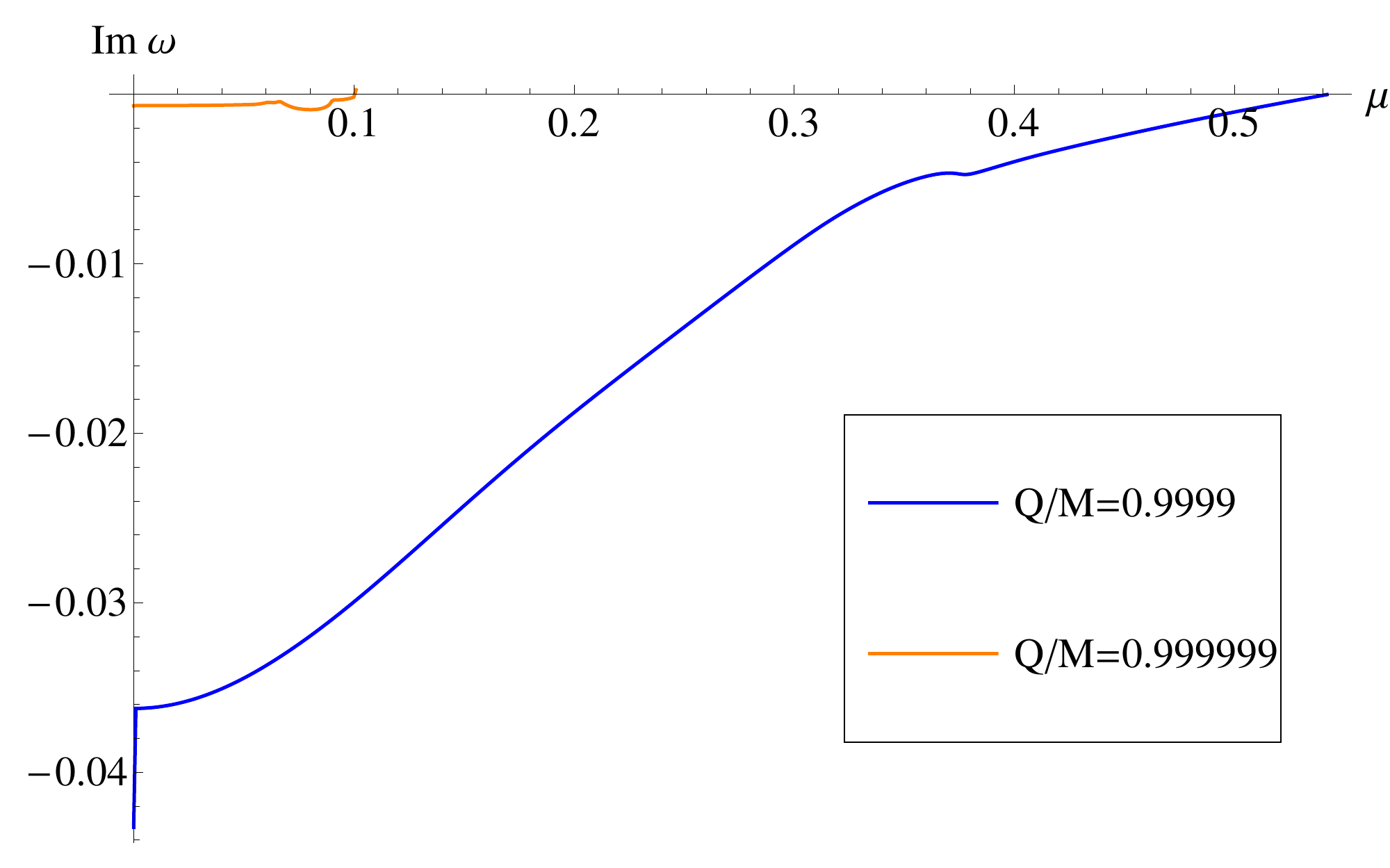}\\
{\scriptsize Figure 3: Dependance of $ \mbox{Re}\,\omega$ on the mass $\mu$ of}& & {\scriptsize Figure 4: Dependance of $ \mbox{Im}\,\omega$ on the mass $\mu$ of}\\
\hspace*{1.3cm}{\scriptsize the scalar field with the charge} & & \hspace*{1.2cm} {\scriptsize the scalar field with the charge}\\
\hspace*{1.3cm}{\scriptsize $q=0.075$, $l=1$.} & & \hspace*{1.4cm}{\scriptsize $q=0.075$, $l=1$.}
\end{tabular} 
\end{center}
It needs to be said that  the calculations leading to results depicted in Figures  $1$-$4$
were carried out\footnote{Note that $l=0$ corresponds to a trivial situation where NC effects disappear.}  for  $l=1$, 
 as well as for the three values of $m,$ namely $m= 0, \pm 1$.
However, as readily seen from figures, the curves corresponding to three different values of $m$ cannot  be distinguished,
actually. Nevertheless, this doesn't  mean that these three curves  coincide  identically.
Just contrary, they are not identical, as can be easily verified    simply by  improving the resolution and
by letting the graphs show a higher level of details.
To that purpose, in the next section we consider the differences of frequences $\omega^{\pm}=\omega_{m=\pm
1}-\omega_{m=0}$, which indeed  appear to be  nonvanishing.
In this regard we observe
that the deformation $a$ and the azimuthal quantum number $m$ always come in pair, implying that the
mode with  $m=0$  actually corresponds to the limit $a \rightarrow 0,$ that is the absence of
a deformation. It is therefore clear that  the differences  $\omega^{\pm}$   encode the effect of a
space-time deformation. This is what  makes them interesting and  worthy of  a separate analysis,
which we leave for the final section. 

Some previous works \cite{ohashi, Konoplya:2004wg}  on QNMs
have studied the decay of the massive scalar field
and found that there exist  perturbations with arbitrary long life when the field mass has special values.  These modes are called  quasiresonances and they are characterized by the zero value of the imaginary part of the frequency. One may ask if such quasiresonances appear in our results. If they do, they should certainly be visible from Figure $4$. Although the graphs on Figure $4$ do not intersect the horizontal axis, it is though clearly seen
that  they approach the zero value as the mass of the field grows. The intersection point itself cannot be seen on Figure $4$ due to the Mathematica's performance. Namely,
when the value of Im\,$\omega$ approaches zero, the increment with which Mathematica  performs calculation, and which corresponds to a numerical error, becomes of the order of the result itself, thus rendering a numerical approximation inadequate. This is why  in our analysis we have been proceeding with  a calculation only up to the point where the Mathematica starts to yield the results that are of  order of a numerical error (and are thus  unreliable). We plan to investigate this issue in more details in future work with the WKB and the continued fraction approach.

The real part of $\omega$ grows linearly with both $q$ and $\mu$, as
expected. The imaginary part reaches the
saturation point for larger $q$, while for larger $\mu$ it tends to zero, signalizing the appearance of quasiresonances.
Moreover, the figures clearly show that the closer is the system to reach
the extremal conditions, the smaller is the value of $q$ at  which the imaginary part  of the
QNMs frequency  saturates.
The $q$ dependance is in
agreement with the numerical results for QNMs of the RN black hole presented in
\cite{QNMRNBrazilci}.

\section{Discussion and final remarks}

We have seen from the solutions
(\ref{Reg1}) and (\ref{Reg2}) that the NC correction only appears in the Region 2, that is when
$x\ll 1$. However, due to the matching procedure, we find a non-zero NC effect on the QNMs spectrum.
The effect can be described as a  Zeeman-like splitting
in the spectrum, manifested by the coupling between the deformation parameter $a$ and the
azimuthal (magnetic) quantum number $m$. In Figures $5$ and $6$ we plot the dependance of  frequency  splitting,
$\omega^\pm = \omega_{m=\pm 1}-\omega_{m=0}$,
of real/imaginary part of $\omega$ on the scalar field charge $q$. The plots are done for
the following set of parameters: $Q/M=0.999999$, $\mu=0.05$ and $l=1$. The green line
represents $\omega^+=\omega_{m=1}-\omega_{m=0}$, while the red line represents
$\omega^-=\omega_{m=-1}-\omega_{m=0}$.

\begin{center}
\begin{tabular}{ccc}
\includegraphics[scale=0.3]{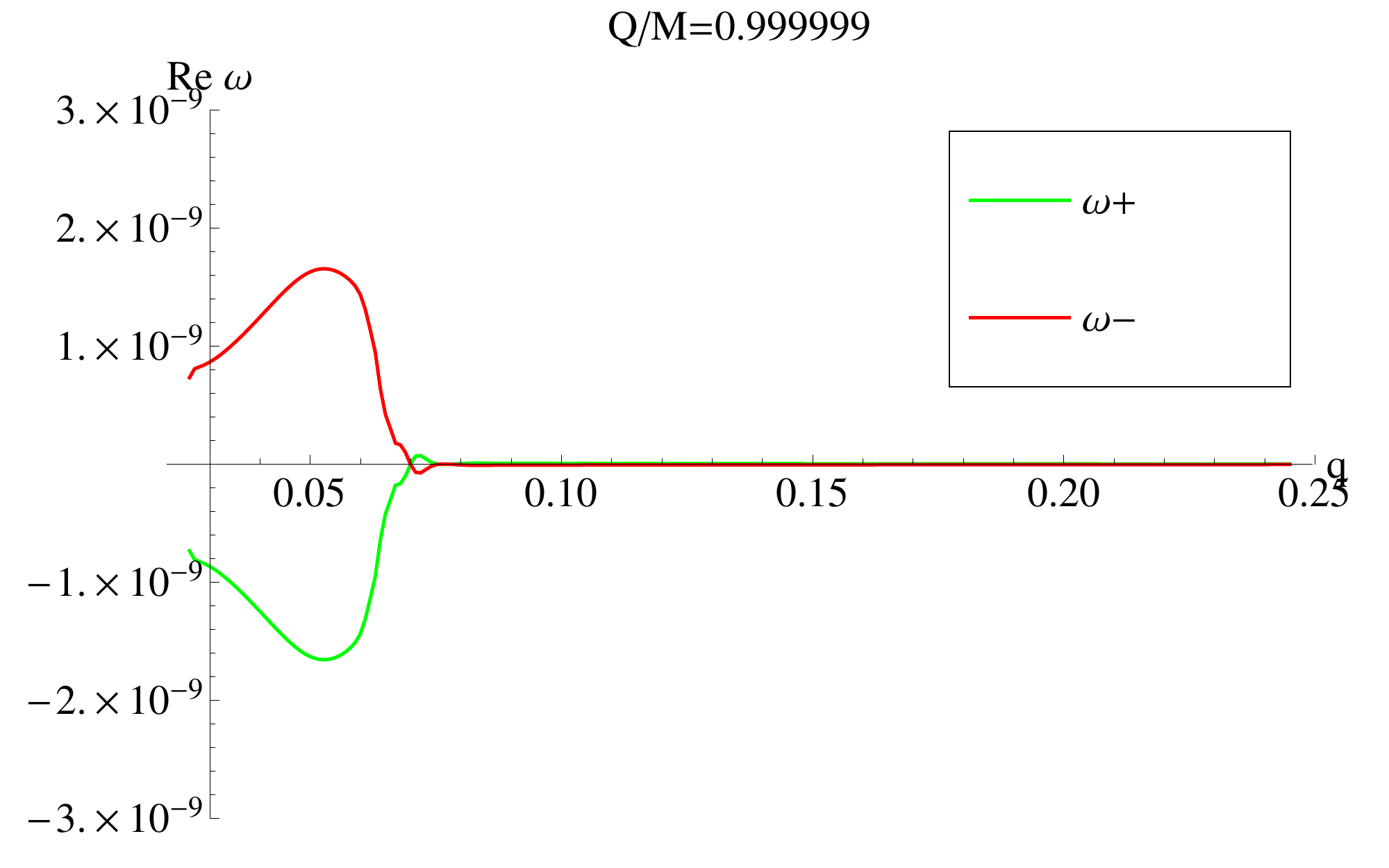}& &
\includegraphics[scale=0.3]{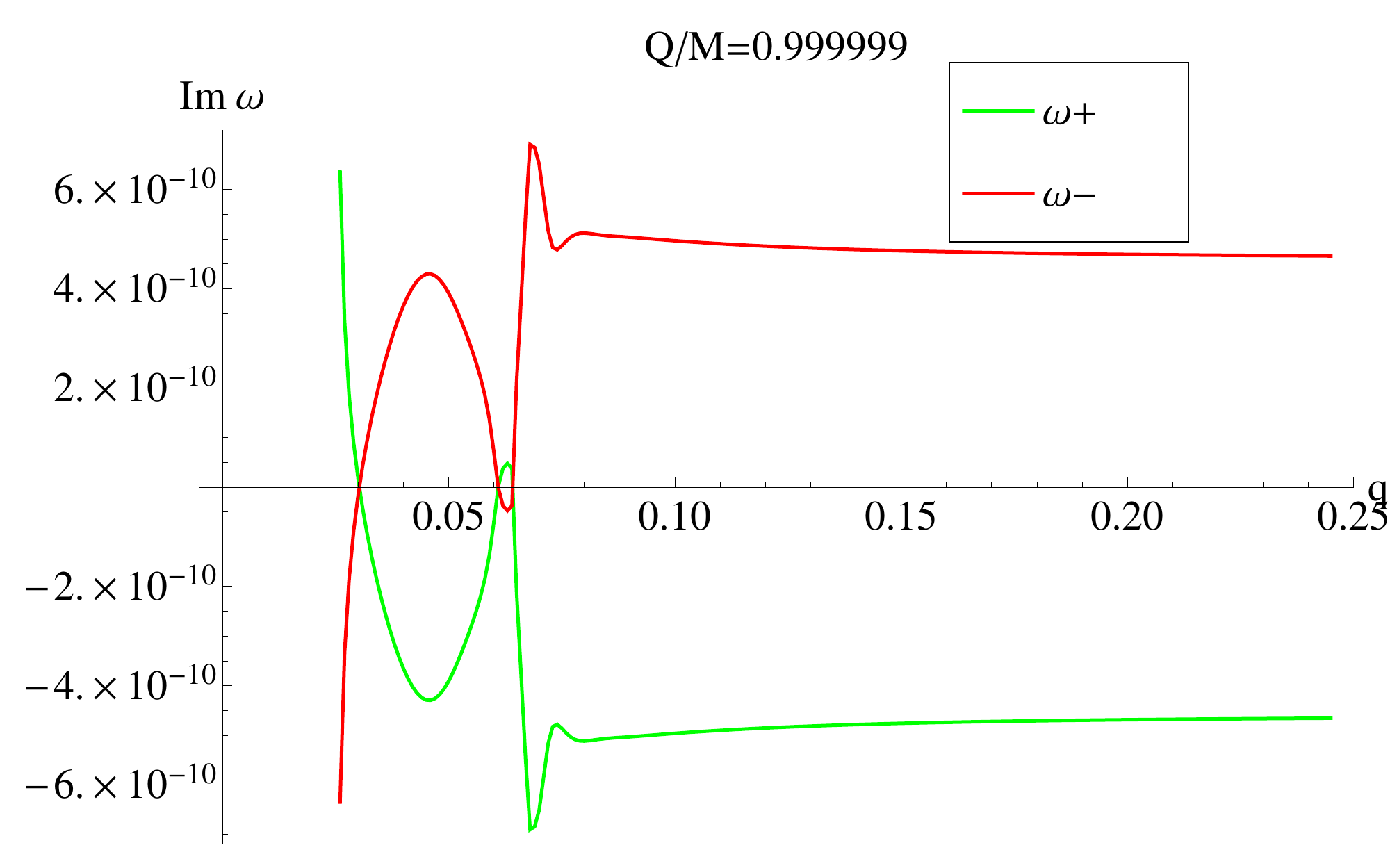}\\
{\scriptsize Figure 5: Dependance of frequency splitting of }& & {\scriptsize Figure 6: Dependance of frequency splitting of }\\
\hspace*{1cm}{\scriptsize $\mbox{Re}\,\omega$ on the charge $q$ of the scalar } & & \hspace*{0.9cm} {\scriptsize $\mbox{Im}\,\omega$ on the charge $q$ of the scalar}\\
\hspace*{1.2cm}{\scriptsize field with the mass $\mu=0.05$, $l=1$. } & &  \hspace*{1.1cm} {\scriptsize field with the mass $\mu=0.05$, $l=1$.}

\end{tabular} 
\end{center}

We also show (see Figures $7$ and $8$) the dependance of  frequency splitting, $\omega^\pm$, on the scalar field mass $\mu$, for the
following set of parameters: $Q/M=0.999999$, $q=0.075$ and $l=1$. Again, the green line
represents $\omega^+=\omega_{m=1}-\omega_{m=0}$, while the red line represents
$\omega^-=\omega_{m=-1}-\omega_{m=0}$.

\begin{center}
\begin{tabular}{ccc}
\includegraphics[scale=0.3]{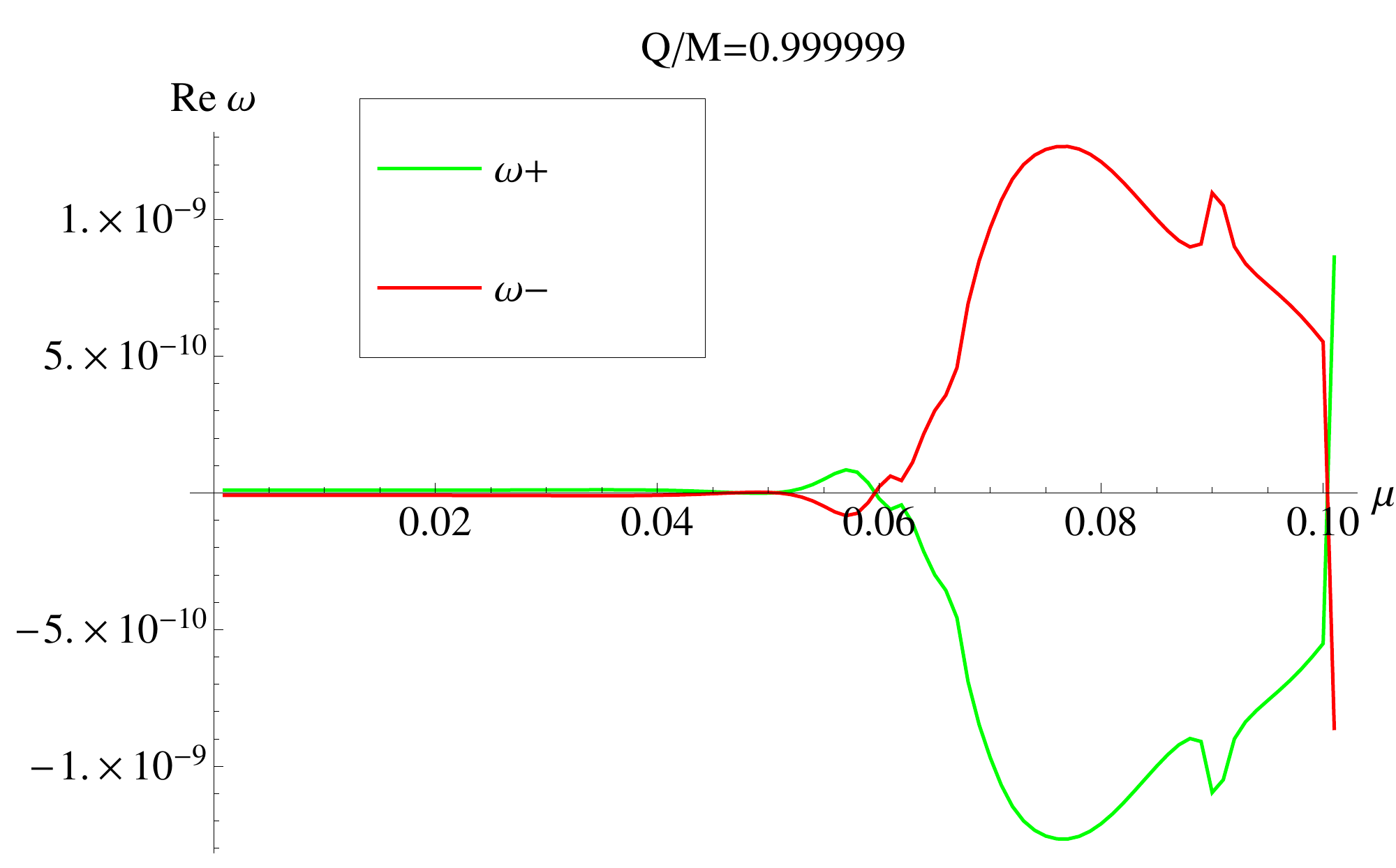}& &
\includegraphics[scale=0.3]{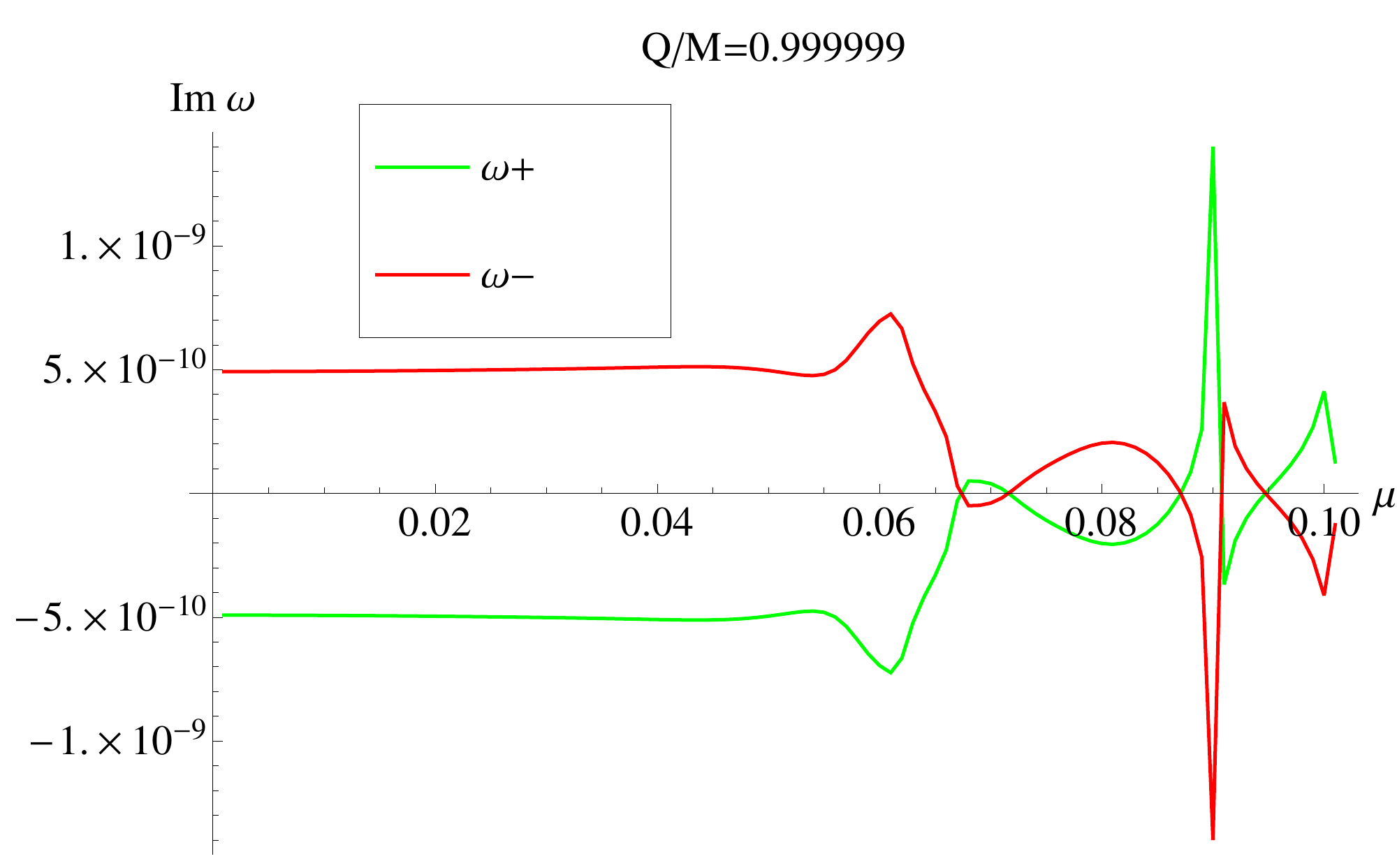}\\
{\scriptsize Figure 7: Dependance of frequency splitting of }& & {\scriptsize Figure 8: Dependance of frequency splitting of }\\
\hspace*{1.3cm}{\scriptsize $\mbox{Re}\,\omega$ on the mass $\mu$ of the scalar field} & & \hspace*{1.2cm} {\scriptsize $\mbox{Im}\,\omega$ on the mass $\mu$ of the scalar field}\\
\hspace*{0.9cm}{\scriptsize  with the charge $q=0.075$, $l=1$. } & &  \hspace*{0.7cm}{\scriptsize  with the charge $q=0.075$, $l=1$.}
\end{tabular} 
\end{center}

The frequency splitting is small, as expected. To have an idea of how small, one can estimate
$\frac{\delta\omega}{\omega}$ for the imaginary part of $\omega$ in the case of $q$ dependance from
Figures $2$ and $6$ and obtain $\frac{\delta\omega}{\omega}\approx 10^{-6}$. However, the effect is very
important qualitatively, since it predicts a Zeeman-like splitting of the QNMs spectrum in the
presence of noncommutativity. 

The phenomenon of frequency splitting can be made even more
evident if we turn our attention to the graphs with $l=2$.
\begin{center}
\begin{tabular}{ccc}
\includegraphics[scale=0.3]{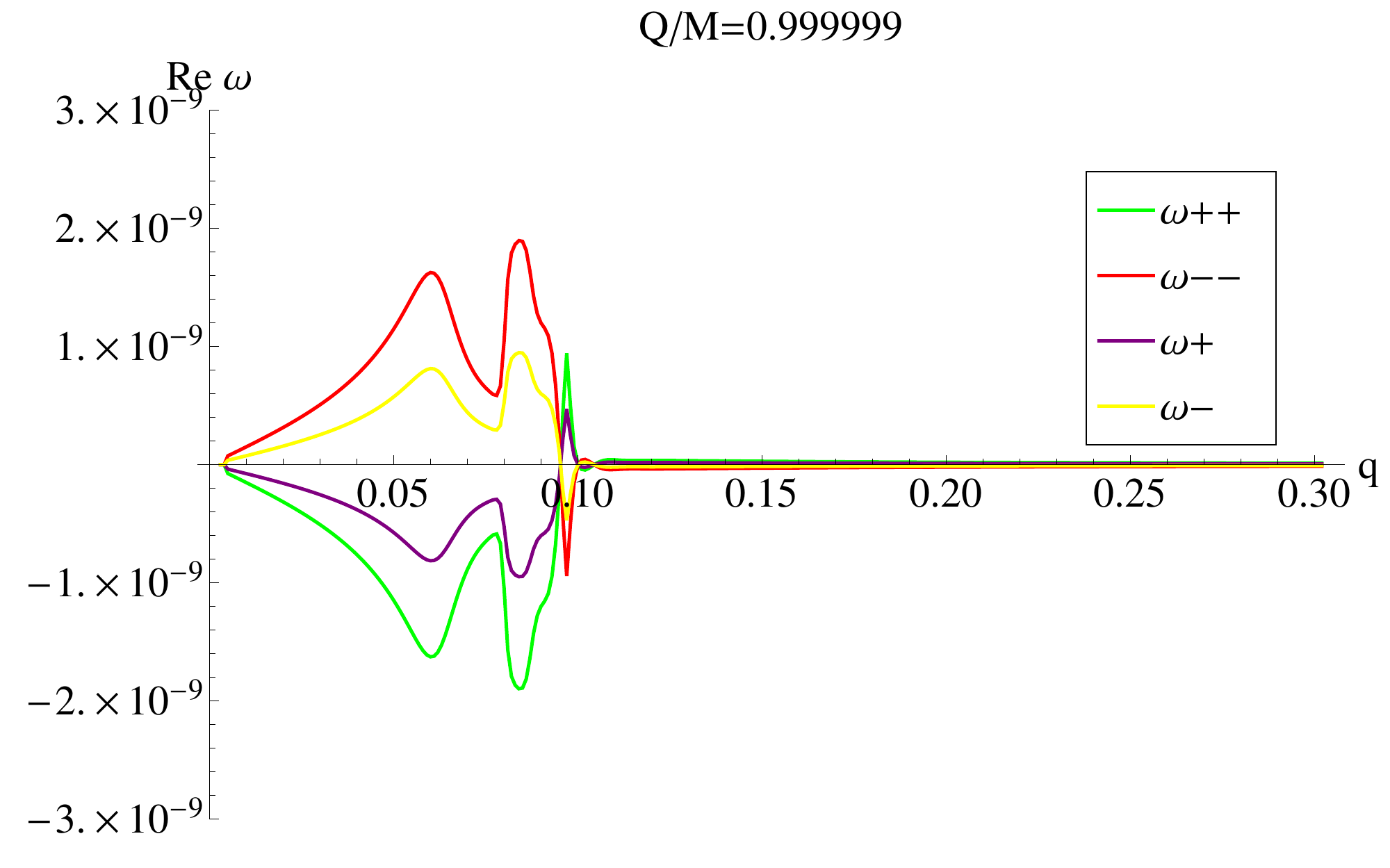}& & \includegraphics[scale=0.3]{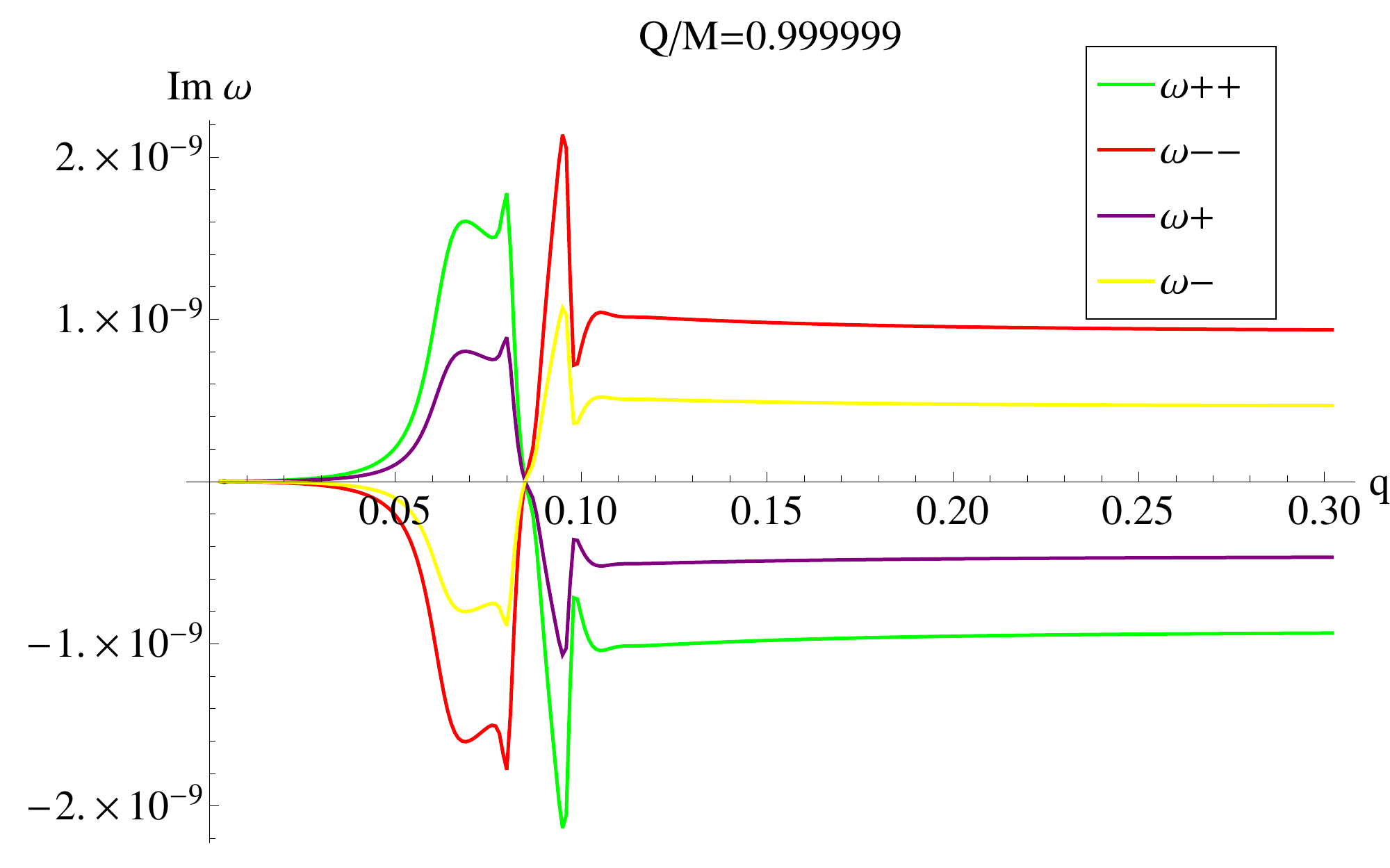}\\
{\scriptsize Figure 9: Dependance of frequency splitting of }& & {\scriptsize Figure 10: Dependance of frequency splitting of }\\
\hspace*{1cm}{\scriptsize $\mbox{Re}\,\omega$ on the charge $q$ of the scalar } & & \hspace*{0.9cm} {\scriptsize $\mbox{Im}\,\omega$ on the charge $q$ of the scalar}\\
\hspace*{1.2cm}{\scriptsize field with the mass $\mu=0.05$, $l=2$. } & &  \hspace*{1.1cm} {\scriptsize field with the mass $\mu=0.05$, $l=2$.}
\end{tabular} 
\end{center}
From Figures $9$ and $10$ we clearly see the frequency splitting, indicating lines that correspond to $m=\pm 2$ and $m=\pm 1$.
Moreover, from all above graphs it is manifest that $~{\omega^{+}} = - {\omega^{-}} ~$ and  $~{\omega^{++}} = - {\omega^{--}}~$, a feature that was expected due  to a parity symmetry.
 Here we use the following notation $\omega^{\pm\pm}=\omega_{m=\pm 2}-\omega_{m=0}$ and $\omega^{\pm}=\omega_{m=\pm 1}-\omega_{m=0}$. For completeness, on Figures $11$ and $12$ we demonstrate the behaviour of the real and imaginary part of the fundamental frequency versus scalar field charge
$q$ for $l=2$ excitations.
\begin{center}
\begin{tabular}{ccc}
\includegraphics[scale=0.3]{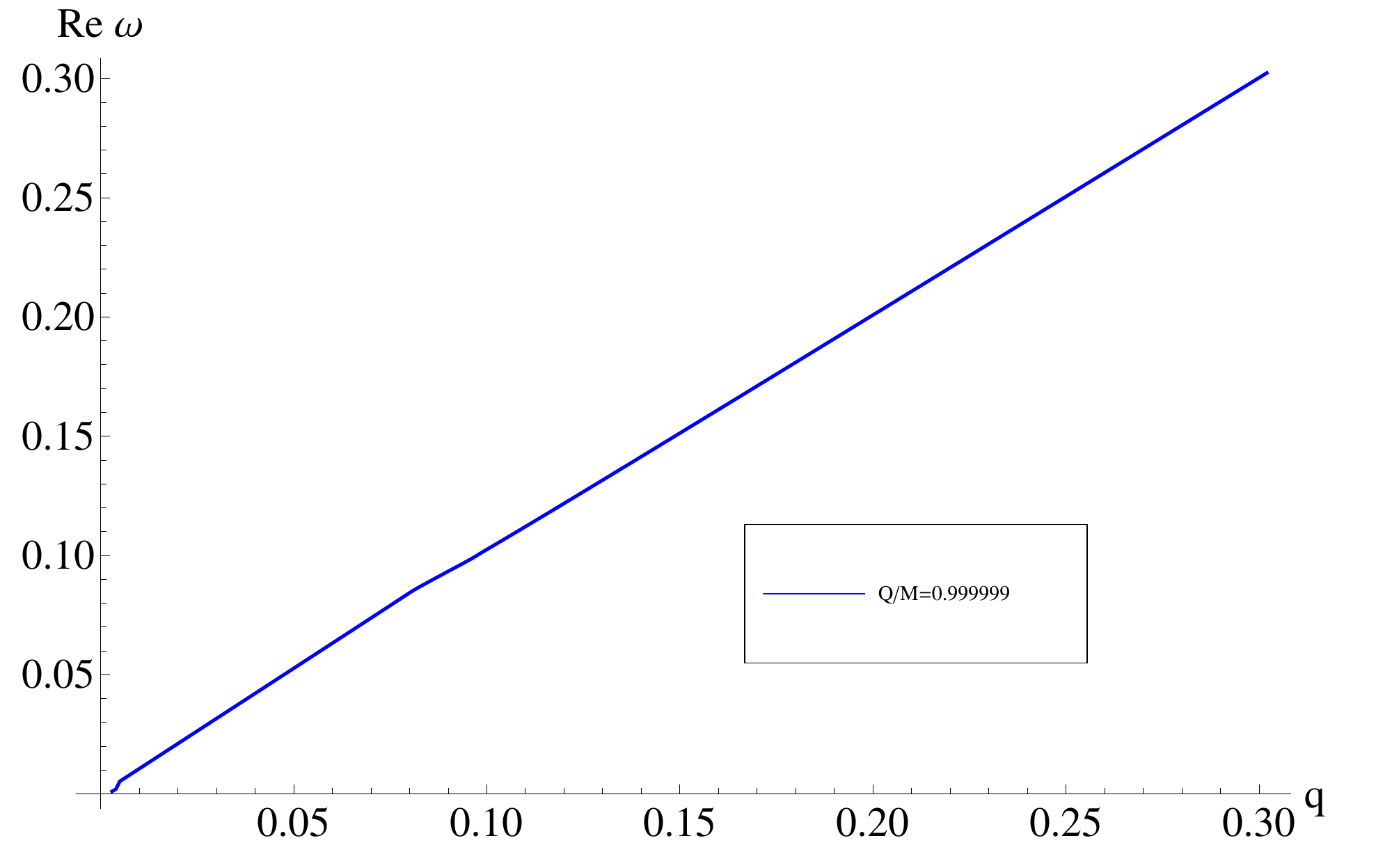}& & \includegraphics[scale=0.3]{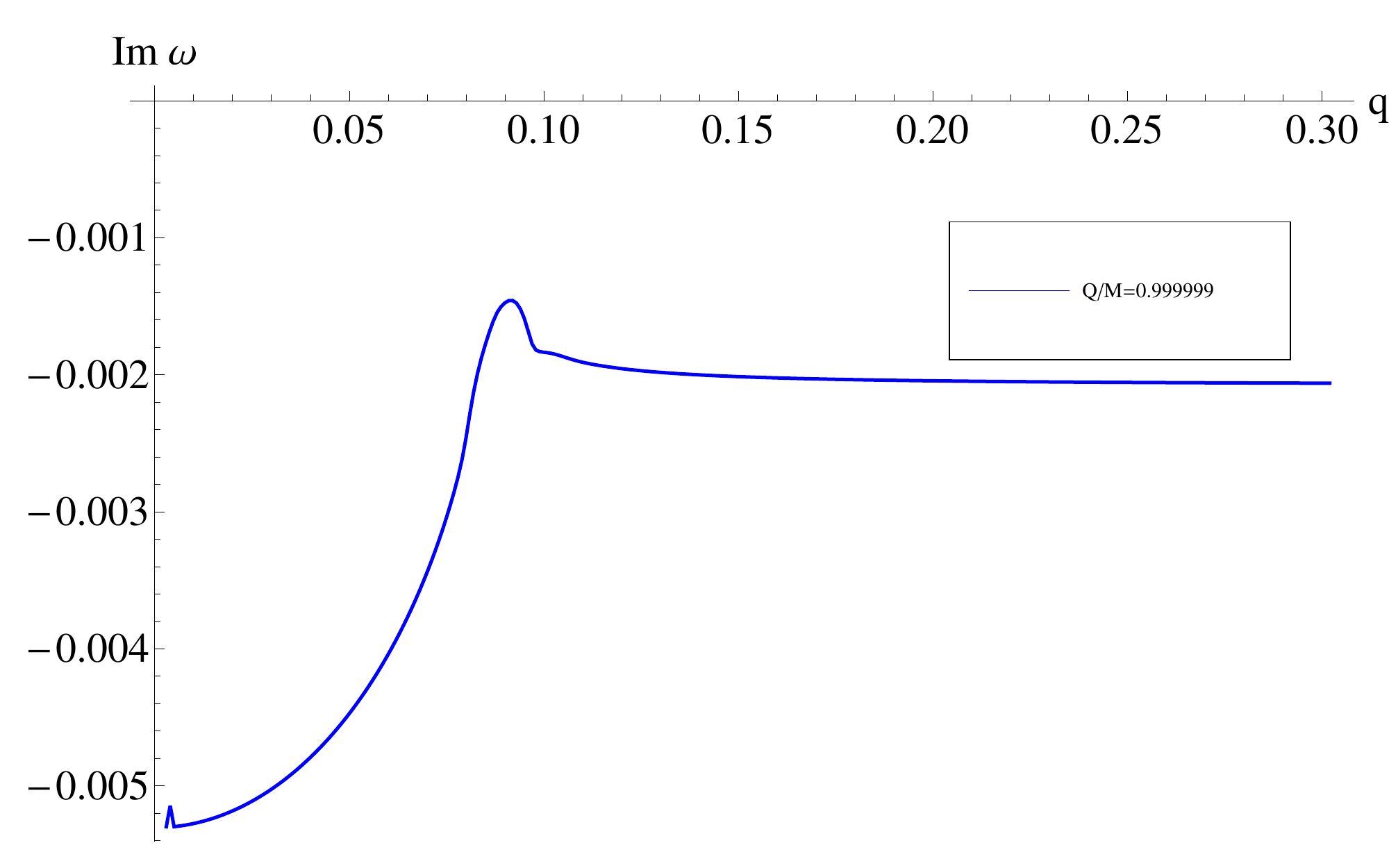}\\
{\scriptsize Figure 11: Dependance of $ \mbox{Re}\,\omega$ on the charge $q$ }& & {\scriptsize Figure 12: Dependance of $ \mbox{Im}\,\omega$ on the charge $q$}\\
\hspace*{1cm}{\scriptsize of the scalar field with the mass } & & \hspace*{0.9cm} {\scriptsize of the scalar field with the mass}\\
\hspace*{-1.3cm}{\scriptsize $\mu=0.05$, $l=2$.} & & \hspace*{-1.2cm}{\scriptsize $\mu=0.05$, $l=2$.}
\end{tabular} 
\end{center}

The frequency splitting
manifests itself as a coupling between the deformation parameter $a$ and the azimuthal
(magnetic) quantum number $m$.
At first glance, a similar behaviour can be found in the QNMs spectrum of 
the Kerr black hole
evaluated in the limit of slow rotation \cite{rew,ferarimashwkb}, where the magnetic quantum number $m$ couples with the black hole angular momentum $J$. Described feature would suggest the existence of a specific  kind of duality  between noncommutative and non-rotating systems on the
one side and standard commutative and rotating systems on the other side. This duality has already been observed in some 
lower dimensional systems \cite{Gupta:2015uga,Gupta:2017lwk,ent}\footnote{Also 
more generally, a duality between generic noncommutative and commutative settings was
claimed, with the gravitational background in the latter case being modified \cite{tureanu}}.
However, a closer inspection  shows that these two spectra are not equivalent or dual to each other, since in the case of rotating black hole in linear approximation, in addition to the term proportional to $m J$, there is another contribution, proportional to $J$ alone, which is nonzero for $m=0$. This is different from the dependence on the noncommutative scale $a$
encountered in our analysis. This shows that a true relationship between these two settings (commutative and noncommutative) still needs to be found and we plan to address this problem in future work.
What we have shown definitely is that the QNMs spectrum of a charged scalar field in the noncommutative RN BH background exhibits a Zeeman-like splitting.

So far we discussed properties of the numerical solution for the QNMs spectrum. Let us briefly
comment on an analytical solution. The analytic solution is possible in a special regime of
the system  parameters: $\tau \ll 1$, $q^2 > G\mu^2$ and $\sigma > 1$. In this case the quantization
condition can be written as
\begin{equation}
\frac{1}{\Gamma(\frac{1}{2} - i\sigma +ik -i \Omega - \tilde{\rho})} = D(\omega, \tau) {\bigg( -2 i
\sqrt{\omega^2 - \mu^2} ~r_+ \tau \bigg) }^{ -2i\sigma}  \ll 1 , \label{AnSol1}
\end{equation}
where the quantity $D(\omega, \tau)$  is defined as
{\small
\begin{eqnarray}
&&D(\omega, \tau) =\label{DOmegaTau}\\
&&-  \frac{ \Gamma(2i\sigma + 2 \tilde{\rho}) \Gamma(1+2i\sigma) \Gamma(\frac{1}{2}
-i\sigma -ik -
\tilde{\rho} ) \Gamma(\frac{1}{2} -i\sigma  - i \kappa  )  \tau^{-2
\tilde{\rho}}}{\Gamma(1-2i\sigma) \Gamma(-2i\sigma - 2 \tilde{\rho}) \Gamma(\frac{1}{2} +i\sigma -ik + \tilde{\rho} ) \Gamma(\frac{1}{2} +i\sigma +ik - i
\Omega +\tilde{\rho} ) \Gamma(\frac{1}{2} +i\sigma -i \kappa)}.   \nonumber
\end{eqnarray}
}
This means that the value of the argument of $\Gamma$-function needs to be very close to any of the
poles, implying
\begin{equation}
\frac{1}{2} - i\sigma +ik -i \Omega - \tilde{\rho} = 
-n + \eta (n) \epsilon + O(\epsilon^2) ,\label{AnSol2}
\end{equation}
where $\epsilon$ is a small parameter \cite{Hod:2010hw}  which scales with the power of $\tau$.
The quantum number $n$ is a nonnegative integer. It takes the role of an overtone number which characterizes the QNMs spectrum.
The mode with $n=0$ is the fundamental mode. In addition, the quantity  $\eta (n)$ can be calculated
\cite{Hod:2010hw, Hod:2017gvn} as 
$\eta (n) = \frac{D(\omega, 0)}{{(-1)}^n n!},$ where  $D(\omega, 0)$ is (\ref{DOmegaTau}) evaluated
in the extremal limit  $Q \rightarrow  \sqrt{G} M$.
The condition (\ref{AnSol2}) directly gives the QNMs spectrum of the RN black hole perturbed by a charged matter
and in a presence of deformed (noncommutative) structure of space-time
{\tiny  
\setlength{\abovedisplayskip}{6pt}
\setlength{\belowdisplayskip}{\abovedisplayskip}
\setlength{\abovedisplayshortskip}{0pt}
\setlength{\belowdisplayshortskip}{3pt}
\begin{equation} 
\begin{split}
& \omega = \frac{qQ}{r_+}  + \frac{\sqrt{G^2 M^2 - GQ^2}}{r_+^2} \Bigg(  qQ - \sqrt{q^2 Q^2 - \mu^2
r_+^2 - {(l+ \frac{1}{2})}^2} - i (n + \frac{1}{2})     + i\eta (n) \epsilon    
+ i \frac{am}{2 r_+^3} \frac{G q Q^3 - GMqQ r_+}{ \sqrt{q^2 Q^2 - \mu^2 r_+^2 - {(l+
\frac{1}{2})}^2}} \Bigg)  \\
& + \frac{G^2 M^2 - GQ^2}{r_+^3}
\Bigg(  \frac{17}{4} qQ - 2\sqrt{q^2 Q^2 - \mu^2 r_+^2 - {(l+ \frac{1}{2})}^2} - 2i (n + \frac{1}{2})   
+  \frac{qQ}{4} \frac{1}{\sqrt{q^2 Q^2 - \mu^2 r_+^2 - {(l+ \frac{1}{2})}^2}} \bigg(9i (n + \frac{1}{2}) - 11qQ \bigg)        \Bigg), 
\end{split}
\label{AnSol3}
\end{equation}}  
where  $~  n= 0,1,2,...$

  Note that to get this formula, we have utilised the fact that the rhs of (\ref{AnSol1}) contains the factor of
$ {(-i)}^{-2i \sigma} = e^{-\pi \sigma}$, which may severely suppress the whole expression as long as the value of $\sigma $ satisfies $\sigma > 1$. We also use the  additional assumption  that the actual  spectrum  slightly  deviates  from its classical value of
$ \omega = qQ/r_+ $.
In this case $\sigma$ is almost real and has enough large real part.
Consequently the right hand side of (\ref{AnSol1}) will come close to zero and so will the left hand side.
This means that the argument of any  $\Gamma$- function in the denominator of the lhs of (\ref{QNMcondition}) might reach the pole and the corresponding  $\Gamma$-function   may consequently blow up,  essentially leading  to  three different branches for the spectrum.
However, not each of these branches of the solution for the spectrum would be physically acceptable, in a sense that not all of them would comply to the requirement that  the actual spectrum needs to be centered around  the classical value of  $ \omega = qQ/r_+ $. This argument singles out the appropriate $\Gamma$-function, which leads to the condition (\ref{AnSol1}) and the spectrum (\ref{AnSol3}).

Although derived under rather stringent conditions, the expression (\ref{AnSol3}) may nevertheless indicate, at
least qualitatively, certain landmark properties of the QNMs spectrum for the problem considered.
In the commutative limit, $a\to 0$, and expanding up to first order in $\frac{\tau}{r_+}$ the
result (\ref{AnSol3}) reduces to the solution found in \cite{Hod:2010hw}
and it can be checked that  by  its general appearance, the structure of this solution is the same as that of the corresponding  result obtained by the WKB method \cite{QNMode}. Moreover, a comparison of our results to those obtained in \cite{QNMRNBrazilci} by the continued fraction method and for the near extremal case ($Q/M$ close to $1$)  shows very similar patterns for  the frequency curves  $\omega$ versus $q$, obtained  in these two approaches.
We postpone the comparation of our results with the results of \cite{Konoplya:2013rxa} for the subsequent paper where we plan to analyze the most general equation of motion (\ref{EoMR})
by using the method of continued fractions and WKB.
Finally, note that, even in the
commutative limit, our analytical result holds up to second order in $\frac{\tau}{r_+}$, while the result of \cite{Hod:2010hw} is only valid up to first order in $\frac{\tau}{r_+}$. Of
course, we again note the appearance of the Zeeman-like splitting
in the spectrum.

Finally, let us briefly discuss the impact of space-time deformation on Hawking
temperature of the RN black hole. In particular, it is of interest to analyse whether it
changes or not. To get the answer to this question, one may use the
approach  based on a semiclassical
method of modeling Hawking radiation as a tunneling effect from the inside to the outside of the horizon \cite{kraus-wilczek,parikh-wilczek}. Up to now we have treated the scalar field $\phi$ as a classical, noncommutative field. In order to study the Hawking radiation, the scalar field $\phi$ has to be quantized. Following \cite{UnitaryNCQFT}, we know that a careful formulation of NC quantum field theory leads to no problems with unitarity and causality.

The method for calculating changes in Hawking temperature  consists of calculating the imaginary part of the classical Hamilton-Jacobi action $S$
which fixes \cite{Kerner,Mitra:2006qa,Li:2008ws} the tunneling
probability amplitude $\Gamma$ through the relation $\Gamma  = \exp ( -\frac{4}{\hbar}
\mbox{Im} S )$. In our analysis  the required quantity is obtained  by the  Hamilton-Jacobi
method \cite{angheben,padmanabhan},
which assumes that the field $\phi$ may be written as
\begin{equation}
\label{final1}
\phi = \exp \Big(-\frac{i}{\hbar} S + O \big( {(-\frac{i}{\hbar})}^2  \big)\Big),
\end{equation}
where in our case $\phi$ satisfies equation (\ref{EomPhiExp1}).
$S$ may be simplified by separating out the  variables in terms of which the system is
described. This is possible due to the symmetry properties of the system, which is made
obvious by the existence of the two Killing vectors. Therefore, one may write $S= -Et + j
\varphi + {\cal R}(r,\theta)$.
Inserting (\ref{final1}) into (\ref{EomPhiExp1}) and using the above separation of $S$,
gives rise to the  equation for ${\cal R}$,
\begin{equation}
\label{final3}
\frac{E^2}{f} + \frac{aqQ}{r^2} f (\partial_r {\cal R}) (\partial_{\varphi} S) -
\frac{1}{r^2} \Xi (\theta, \varphi) = f {(\partial_r {\cal R})}^2.
\end{equation}
Here $\Xi (\theta, \varphi)$ is the abbreviation for
\begin{equation} 
\label{final4}
\Xi (\theta, \varphi)  \equiv -\frac{1}{\hbar^2} \Big(  {(\partial_{\theta} S)}^2 + i \hbar
\partial_{\theta}^2 S + i\hbar \cot \theta  \partial_{\theta} S + \frac{1}{\sin^2\theta} 
{(\partial_{\varphi} S)}^2 + i \hbar \frac{1}{\sin^2 \theta}  \partial_{\varphi}^2 S  \Big).
\end{equation}
Solving (\ref{final3}) with respect to $\partial_r {\cal R}$ gives
\begin{equation} 
\label{final5}
\partial_r {\cal R} =  \frac{aqQ}{2 r^2} ~ \partial_{\varphi} S \pm
\sqrt{\big(  \frac{E^2}{f^2} - \frac{1}{r^2 f}  \Xi (\theta, \varphi)    \big) 
\Big( 1+  \frac{a^2 q^2 Q^2}{4 r^4} \frac{{(\partial_{\varphi} S)}^2}{ \frac{E^2}{f^2}
-\frac{1}{r^2 f}  \Xi (\theta, \varphi) } \Big) }.
\end{equation}
Expanding with respect to the deformation parameter $a$ and keeping terms up to order $O(a^2)$
yields
\begin{equation}
\label{final6}
{\cal R}(r) = \int_{r_{in}}^{r} dr' \Big(\frac{aqQ}{2 r'^2} \partial_{\varphi} S \pm  
\frac{1}{f} \sqrt{E^2 - \frac{f}{r'^2} \Xi (\theta, \varphi) } \big( 1+
\frac{f^2}{8r'^4}\frac{a^2 q^2 Q^2  {(\partial_{\varphi} S)}^2}{E^2 - \frac{f}{r'^2} \Xi
(\theta, \varphi)} \big) \Big),
\end{equation}
with $r_{-} < r_{in} < r_+ $ and $r > r_+$ . What is needed is the imaginary part of the
classical action $S$ (and therefore of the function ${\cal R}$). This can only come from the
integration over the contour encircling the pole of the integrand. Since the pole is located at the
zero of $f$ (outer horizon), the calculation of the
corresponding residue then clearly shows that the terms scaling with $f$ and $f^2$ simply 
disappear. Therefore, the noncommutativity in space-time does not affect the Hawking temperature, it
is shifted  neither above nor below the standard Hawking temperature for the RN black hole. 

This result is expected, since in our model the gravitational field (geometry) is not affected
by the noncommutativity. It would be interesting to analyse effects of the NC scalar fields to the
background geometry. Some results in this direction are presented in \cite{Nicolini2005,
Kobakhidze2007}, but they only concern Schwarzschild black hole background. The backreaction of the
NC scalar field on the geometry could introduce (among other effects) a shift of the horizon, leading to
a change of the Hawking temperature. This analysis we postpone for future work.

In addition to the analysis we performed here, one can straightforwardly include NC spinor and NC
vector fields and calculate the QNMs spectrum of these fields. A more difficult task is to analyse
the gravitational QNMs. For that one needs an action describing a NC gravitation field. There are
various suggestions in the literature \cite{NCgravity}. The main difficulty is certainly the fact
that the first order correction in the deformation parameter vanishes. Therefore, the first
non-vanishing correction to the NC gravitational action is second order in the deformation
parameter, leading to more cumbersome calculations. All these effects we plan to investigate in our
future work.

\vskip1cm \noindent 
{\bf Acknowledgement}
\hskip0.3cm
We would like to thank Tajron Juri\'c, Svetislav Mijatovi\' c, Danijel Pikuti\'c and Voja
Radovanovi\'c for
fruitful discussion and useful comments. The work of M.D.C. and N.K.
is
supported by project
ON171031 of the Serbian Ministry of Education and Science. The work of A.S. is  supported by
Croatian Science Foundation under the project (IP-2014-09-9582) and it is partially supported by the
H2020 CSA Twinning project No. 692194, RBI-T-WINNING. This work is partially
supported by ICTP-SEENET-MTP Project
PRJ09 "Cosmology and Strings" in frame of the Southeastern European Network in
Theoretical and Mathematical Physics. and by the Action MP1405 QSPACE from the European 
Cooperation in  Science
and  Technology  (COST).

\appendix

\renewcommand{\theequation}{\Alph{section}.\arabic{equation}}
\initiate
\section{Twisted differential calculus}

Space-time symmetries such as diffeomorphism symmetry of GR, or its subgroups Poincar\'e
symmetry or conformal symmetry, are generated by vector fields. The Lie algebra of vector fields we
label by $\Xi$ and its universal enveloping algebra by $U\Xi$. Note that $U\Xi$ is a Hopf algebra.

A well defined way to deform the symmetry Hopf algebra is via a Drinfeld twist \cite{Drinfeld}. The
twist $\mathcal{F}$ is an invertible element of $U\Xi \otimes U\Xi $
satisfying the following properties:
\begin{enumerate}
\item the cocycle condition 
\begin{equation}
(\mathcal{F}\otimes 1)(\Delta\otimes id)\mathcal{F}=(1\otimes \mathcal{F}%
)(id\otimes \Delta)\mathcal{F},  \label{Twcond1}
\end{equation}

\item normalization 
\begin{equation}
(id\otimes \epsilon)\mathcal{F} = (\epsilon\otimes id)\mathcal{F}=1\otimes 1,
\label{Twcond2}
\end{equation}

\item perturbative expansion 
\begin{equation}
\mathcal{F} = 1\otimes 1 + \mathcal{O}(\lambda),  \label{Twcond3}
\end{equation}
\end{enumerate}
where $\lambda $ is a deformation parameter. The last property insures that in the limit
$\lambda\to 0$ the undeformed algebra $U\Xi$ is restored. We shall frequently use the notation (sum
over $\alpha
=1,2,...\infty $ is understood) 
\begin{equation}
\mathcal{F}=\mathrm{f}^{\alpha }\otimes \mathrm{f}_{\alpha },\quad \mathcal{F%
}^{-1}=\bar{\mathrm{f}}^{\alpha }\otimes \bar{\mathrm{f}}_{\alpha },
\label{Fff}
\end{equation}
where, for each value of $\alpha $, $\bar{\mathrm{f}}^{\alpha }$ and $\bar{%
\mathrm{f}}_{\alpha }$ are two distinct elements of $U\Xi $ (and similarly $%
\mathrm{f}^{\alpha }$ and $\mathrm{f}_{\alpha }$ are in $U\Xi
$). The twist
acts on the symmetry Hopf algebra and gives the twisted symmetry (as
deformed Hopf algebra) 
\begin{eqnarray}
\lbrack \xi ,\eta ] &=&(\xi ^{\mu }\partial _{\mu }\eta ^{\rho }-\eta ^{\mu
}\partial _{\mu }\xi ^{\rho })\partial _{\rho },  \notag \\
\Delta^{\mathcal{F}}(\xi ) &=&\mathcal{F}\Delta (\xi )\mathcal{F}^{-1} 
\notag \\
\varepsilon (\xi) &=&0,\quad S^{\mathcal{F}}(\xi )=\mathrm{f}^{\alpha
}S(\mathrm{f}_{\alpha })S(\xi )S(\bar{\mathrm{f}}^{\beta })\bar{\mathrm{f}}
_{\beta }.  \label{TwistedUg}
\end{eqnarray}
Here $\xi, \eta$ are vector fields belonging to $\Xi$. We see that after the twist
deformation, the algebra remains the same, while in general the comultiplication (coproduct)
and
antipode change.  The whole deformation depends on formal parameters which
control classical limit. Twisted (deformed) comultiplication leads to
the deformed Leibnitz rule for the symmetry transformations when acting on
product of fields. 

The twist can be used to deform the commutative geometry of space-time
(vector fields, 1-forms, exterior algebra of forms, tensor algebra) in a well defined way.
More details on the twisted differential geometry can be found in \cite{PMKLWbook}.

We work with an Abelian twist in $D=4$
\begin{equation}
\mathcal{F}=e^{-\frac{i}{2}\theta ^{AB}X_{A}\otimes X_{B}}.  \label{AbelianTwist}
\end{equation}
Here $\theta ^{AB}$ is a constant antisymmetric matrix and indices $A, B =1,\dots, p$ with $p\leq
4$. $X_A$ are commuting vector fields $[X_A,X_B]=0$. This requirement ensures that the cocycle
condition is fulfilled.

Applying the inverse of the twist (\ref{AbelianTwist}) to the usual point-wise
multiplication of functions, $\mu (f\otimes g)=f\cdot g$; $f,g\in A$, we obtain
the $
\star $-product of functions 
\begin{eqnarray}
f\star g &=&\mu \mathcal{F}^{-1}(f\otimes g)  \notag \\
&=&\bar{\mathrm{f}}^{\alpha }(f)\bar{\mathrm{f}}_{\alpha }(g).
\label{FunctionsStar}
\end{eqnarray}
The action of the twist ($\bar{\mathrm{f}}^{\alpha }$ and $\bar{\mathrm{f}}
_{\alpha }$) on the functions $f$ and $g$ is via the Lie derivative.

The product between functions and one-forms is
defined as 
\begin{equation}
h\star \omega =\bar{\mathrm{f}}^{\alpha }(h)\bar{\mathrm{f}}_{\alpha
}(\omega ) \label{FunctionStarForms}
\end{equation}
with a function $h$ and an arbitrary 1-form $\omega $. The action of $\bar{\mathrm{f}}_{\alpha }
$ on forms is given via the Lie derivative. We  often use the
Cartan's formula for the Lie derivative along the vector field $\xi $ of an
arbitrary form $\omega $ 
\begin{equation}
{\mathcal{L}}_{\xi }\omega =\mathrm{d}i_{\xi }\omega +i_{\xi }\mathrm{d}\omega .
\label{CartanF-la}
\end{equation}
Here $\mathrm{d}$ is the exterior derivative and $i_{\xi }$ is the
contraction along the vector field $\xi$. 

Arbitrary forms form an exterior algebra with the wedge product. The
$\star$-wedge product on two arbitrary forms $\omega$ and $\omega^{\prime }$
is 
\begin{equation}
\omega\wedge_\star\omega^{\prime }= \bar{\mathrm{f}}^\alpha(\omega) \wedge 
\bar{\mathrm{f}}_\alpha(\omega^{\prime }).  \label{WedgeStar}
\end{equation}

The usual (commutative) exterior derivative satisfies: 
\begin{eqnarray}
\mathrm{d} (f\star g) &=& \mathrm{d}f\star g + f\star \mathrm{d}g,  \notag \\
\mathrm{d}^2 &=& 0,  \notag \\
\mathrm{d} f &=& (\partial_\mu f) \mathrm{d}x^\mu = (\partial^\star_\mu f)
\star \mathrm{d}x^\mu.  \label{Differential}
\end{eqnarray}
The first property if fulfilled because the usual exterior derivative
commutes with the Lie derivative which enters in the definition of the $\star
$-product. Therefore, we will use the usual exterior derivative as the
noncommutative exterior derivative. Note that the last line of
(\ref{Differential}) gives a definition of the $\partial^\star_\mu$
derivatives.

The usual integral is cyclic under the $\star $-exterior products of forms 
\begin{equation}
\int \omega _{1}\wedge _{\star }\omega _{2}=(-1)^{d_{1}\cdot d_{2}}\int
\omega _{2}\wedge _{\star }\omega _{1},  \label{IntCycl}
\end{equation}
where $d=deg(\omega )$, $d_{1}+d_{2}=4$ provided that $S^{\cal
F}(\bar{\mathrm{f}}
^{\alpha })\bar{\mathrm{f}}_{\alpha }=1$ holds. One can check that
this indeed holds for any Abelian twist. The property (\ref{IntCycl}) can be generalized to 
\begin{equation}
\int \omega _{1}\wedge _{\star }\dots \wedge _{\star }\omega
_{p}=(-1)^{d_{1}\cdot d_{2}\dots \cdot d_{p}}\int \omega _{p}\wedge _{\star
}\omega _{1}\wedge _{\star }\dots \wedge _{\star }\omega _{p-1},  \label{IntCycl2}
\end{equation}
with $d_{1}+d_{2}+\dots +d_{p}=4$. We say that the integral is cyclic. This property is very
important for
construction of NC gauge theories.


\end{document}